\documentclass[journal]{vgtc}                     

\onlineid{3721}

\vgtccategory{Research}

\title{MedVA: An End-to-End Neuro-Symbolic Agentic System \\for Medical Volume Visualization}

\author{%
  Haill An,
  Suhyeon Kim,
  Minjun Kang,
  Eunwoo Lee,
  Bin Sheng,
  Lei Bi$^{\dagger}$, and 
  Younhyun Jung$^{\dagger}$
}

\authorfooter{
  \item
    \textsuperscript{\(\dagger\)} Corresponding authors.

  \item
    H. An, S. Kim, M. Kang, E. Lee, and Y. Jung are with the
    School of Computing, Gachon University, Republic of Korea.
    E-mail: \{xenotic, kih629, alswns1119, ee1225, younhyun.jung\}@gachon.ac.kr.

  \item
    B. Sheng is with the Department of Computer Science and Engineering,
    Shanghai Jiao Tong University, Shanghai, China.
    L. Bi is with the Institute of Translational Medicine,
    National Center for Translational Medicine,
    Shanghai Jiao Tong University, Minhang, Shanghai, China.
    E-mail: \{shengbin, lei.bi\}@sjtu.edu.cn.
}

\abstract{%
  Medical volume visualization requires selecting regions of interest (ROIs) and carefully controlling their relative visual emphasis according to a given clinical intent. Implementing these decisions in conventional workflows demands substantial clinical and visualization expertise and often involves trial-and-error optimization. Recent agentic systems have introduced natural-language interaction and autonomous visualization operations but largely rely on MLLM-based inference throughout the workflow. Although MLLMs encode broad medical knowledge and provide strong reasoning capabilities, such inference may be suboptimal for medical volume visualization, potentially leading to clinically incomplete interpretations of user requests and unreliable ROI identification and visualization optimization. In this work, we present MedVA, an end-to-end neuro-symbolic agentic system for medical volume visualization that addresses these limitations through three complementary agents. The neuro-symbolic intent formulation agent refines MLLM-based interpretations of natural-language requests through symbolic reasoning over established clinical knowledge, which provides more complete, clinically grounded ROI specifications than MLLM-only reasoning. To the best of our knowledge, MedVA introduces the first neuro-symbolic design for agentic volume visualization. The multi-model ROI identification agent directly identifies semantically specified ROIs in the original volume by leveraging complementary large-scale pretrained medical segmentation models. The objective-driven visualization optimization agent explicitly evaluates ROI visibility and occlusion in the original volume using a volume-based visibility objective. Extensive agent-level and system-level evaluations across diverse medical datasets and interaction scenarios support the effectiveness of the individual agents and the consistency of the end-to-end workflow. A formative user study further indicates high usability and practical value among users with different levels of expertise.
}


\teaser{
  \centering
  \includegraphics[width=\linewidth, alt={A view of clouds with orange sunrays shining through from behind.}]{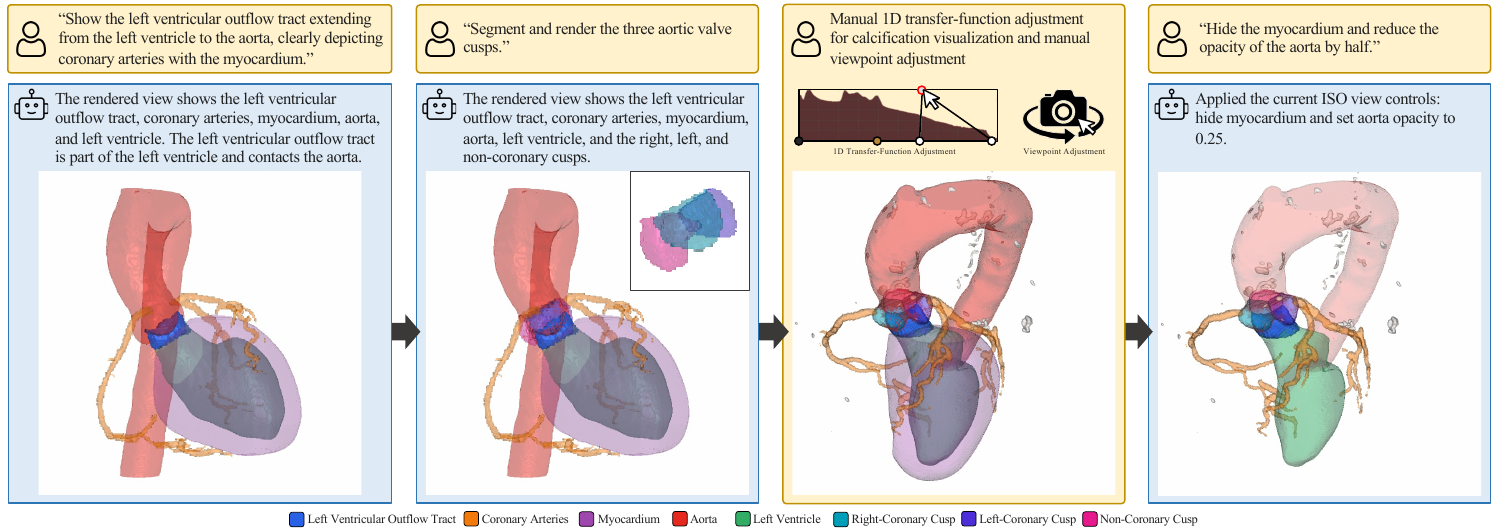}
  \caption{%
  	Representative multi-step interaction with MedVA for cardiac volume visualization. MedVA translates natural-language requests into clinically grounded ROI specifications, identifies multiple ROIs directly in the original volume on demand, and refines ROI-guided visualizations according to their relative priorities in response to subsequent requests. %
  }
  \label{fig:fig1}
}

\graphicspath{{figs/}{figures/}{pictures/}{images/}{./}} 

\usepackage{booktabs}                  
\usepackage{lipsum}                    
\usepackage{mwe}                       
\usepackage{ccicons}                   

\usepackage{booktabs}
\usepackage{tabularx}
\usepackage{array}
\usepackage{makecell}
\usepackage{multirow}
\usepackage[normalem]{ulem}
\usepackage{amssymb} 
\usepackage[table]{xcolor}
\newcolumntype{Y}{>{\raggedright\arraybackslash}X}

\usepackage{mathptmx}                  

\begin{document}


\firstsection{Introduction}

\maketitle
Volume visualization techniques, including direct volume rendering (DVR) and iso-surface rendering (ISR), enable interactive three-dimensional exploration of anatomy and pathology from medical imaging data \cite{Zhang2011VolumeVisualizationTechnical,Zhou2022ReviewThreeDimensional}. Clinically informative volume visualization requires a careful balance between clearly highlighting primary regions of interest (ROIs) and preserving relevant context ROIs. This balance is important for assessing ROI location, extent, and spatial relationships and for supporting diagnosis, treatment planning, and other complex clinical decisions \cite{Svakhine2009IllustrationInspiredDepth,Viola2004ImportanceDrivenVolume,Li2025AttentionDrivenVisual}. Generating an effective volume visualization involves two interdependent decisions: which ROIs should be included for a given clinical intent, and how their visibility should be adjusted according to their relative priorities. Conventional workflows require users to translate these decisions into low-level visualization operations through repeated adjustment of transfer functions, viewpoints, clipping planes, and other rendering parameters. This process demands substantial clinical and visualization expertise, often involves trial-and-error optimization, and may not consistently produce the desired volume visualization \cite{Cai2013AutomaticTransferFunction,RezkSalama2006HighLevelUser}. These challenges remain a major barrier to the broader use of advanced volume visualization in clinical practice.

Recent advances in multimodal large language models (MLLMs) and artificial intelligence (AI) agents have opened new opportunities for more autonomous and intuitive interaction with volume visualization\cite{Liu2024AVATowardsAutonomous,Ai2026NLI4VolVisNaturalLanguage,Mallick2024ChatVisAutomatingScientific,Wang2025IntuiTFMLLMGuided,Sun2026SASAVSelfDirected,Ai2026HiLSVADesignEvaluation,Biswas2026VizGenieToward}. MLLMs support the expression and semantic interpretation of visualization requests in natural language, while their visual perception capabilities allow them to inspect visualization outputs, assess their alignment with user intent, and guide subsequent refinement. By combining MLLMs with planning, tool use, and action execution, AI agents can translate a natural-language request into executable visualization tasks and coordinate the corresponding operations across the volume visualization workflow\cite{Ai2026NLI4VolVisNaturalLanguage,Mallick2024ChatVisAutomatingScientific}. Prior studies have demonstrated these capabilities in volume visualization, including generating visualization scripts, identifying requested ROIs, and manipulating and refining visualization outputs\cite{Liu2024AVATowardsAutonomous,Ai2026NLI4VolVisNaturalLanguage,Mallick2024ChatVisAutomatingScientific,Wang2025IntuiTFMLLMGuided}.

Despite these advances, prior agentic systems still face fundamental limitations when applied to medical volume visualization. A major limitation lies in formulating clinically meaningful visualization goals from potentially underspecified user requests. Such goals need to specify relevant ROIs, their relationships, and relative visualization priorities according to the user intent. Prior agentic systems largely rely on the broad medical knowledge encoded in MLLMs and their general reasoning capabilities, sometimes supplemented by generic knowledge retrieval, to perform this formulation\cite{Ai2026NLI4VolVisNaturalLanguage,Biswas2026VizGenieToward,Sun2026SASAVSelfDirected}. Consequently, the resulting visualization goal may yield a clinically incomplete ROI set, with broad clinical concepts left unresolved or relevant ROIs omitted. ROI identification presents another important limitation. In prior agentic systems, MLLMs commonly infer the requested ROIs from multi-view renderings generated using transfer function presets, selected iso-values, or pre-existing segmentation\cite{Liu2024AVATowardsAutonomous,Ai2026NLI4VolVisNaturalLanguage,Sun2026SASAVSelfDirected}. Such indirect identification can be unreliable when multiple ROIs overlap or exhibit similar visual characteristics in rendered views. Dependence on precomputed renderings can also constrain on-demand identification because newly requested ROIs may require additional preprocessing. Visualization optimization presents a further limitation. Prior agentic systems largely rely on MLLM-based visualization–perception–action (VPA) loops, in which MLLMs inspect intermediate rendering outputs to infer ROI visibility and occlusion and determine subsequent parameter adjustments\cite{Liu2024AVATowardsAutonomous,Ai2026NLI4VolVisNaturalLanguage,Wang2025IntuiTFMLLMGuided}. Such rendering-based implicit optimization can be unreliable when multiple ROIs exhibit complex visibility and occlusion relationships. This VPA loop may also require multiple iterations of MLLM-based evaluation to reach the desired volume visualization, which can substantially reduce optimization efficiency. 

In this work, we propose MedVA, an end-to-end neuro-symbolic agentic system for medical volume visualization. MedVA complements MLLM-based reasoning with explicit clinical knowledge and establishes a clinically grounded pathway from user requests to volume visualizations. MedVA comprises three complementary agents for clinically grounded intent formulation, direct ROI identification in the original volume, and objective-driven visualization optimization. 

Our neuro-symbolic intent formulation agent is designed to derive clinically meaningful visualization goals from potentially underspecified natural-language requests. Neuro-symbolic AI has emerged as a promising paradigm that integrates the flexible semantic inference of neural models with explicit knowledge representations and rule-based reasoning that can constrain and refine uncertain neural interpretations\cite{Hitzler2022NeuroSymbolicApproaches,Wang2025TowardsDataKnowledgeDriven}. Although this complementary approach has been increasingly explored in both AI agents and clinical applications\cite{Hakim2026NeuroSymbolicAgentic,Jia2025medIKALIntegratingKnowledge,Prenosil2025NeuroSymbolicAIAuditable}, it remains largely unexplored in agentic volume visualization. We introduce a neuro-symbolic design for agentic volume visualization. In MedVA, the agent validates and refines MLLM-derived intent interpretation through symbolic reasoning over established clinical knowledge and formulates the resulting visualization goals as an explicit visualization specification. Compared with the MLLM-only baseline, the agent provides a more complete set of clinically grounded ROIs, together with their clinical relationships and relative visualization priorities. This specification also serves as an explicit and verifiable reference for downstream ROI identification and visualization optimization. 

Our multi-model ROI identification agent then directly identifies semantically specified ROIs in the original volume. This capability builds on recent advances in large-scale pretrained medical segmentation models\cite{Wasserthal2023TotalSegmentatorRobust,He2025VISTA3DUnifiedSegmentation,Rokuss2026VoxTellFreeText}, which provide increasingly broad and complementary coverage of diverse anatomical and pathological structures. The agent leverages these advances by using multiple pretrained segmentation models according to the visualization specification and integrating their complementary outputs through deterministic selection and MLLM-based validation. Compared with indirect MLLM-based inference from multi-view renderings, the agent enables more reliable ROI identification while supporting on-demand identification for varying user requests. The resulting ROI masks provide explicit volume-based representations for subsequent visualization optimization. 

Our objective-driven visualization optimization agent translates the identified ROIs and their visualization priorities into final volume visualizations. The agent uses a volume-based visibility objective to explicitly evaluate ROI visibility and occlusion in the original volume. Within the VPA loop, the agent applies this objective to assess the current rendering against the intended visual emphasis and inform parameter adjustments. Compared with implicit MLLM-based visibility assessment of renderings, the agent enables more reliable ROI-guided volume visualization. It can also reduce optimization iterations and improve efficiency. 

We extensively evaluate MedVA through agent-level quantitative and qualitative analyses, system-level evaluations across diverse medical datasets and case studies involving varying user requests and visualization interactions. A formative user study with participants at different levels of expertise further assesses its effectiveness, efficiency, and usability. The main contributions of our work are as follows:

\begin{itemize}
    \item An End-to-End Agentic System for Medical Volume Visualization: We present MedVA, which incorporates explicit clinical knowledge to complement MLLM-based reasoning. The system establishes a clinically grounded pathway from natural-language user requests to ROI-guided volume visualizations through neuro-symbolic intent formulation, direct ROI identification, and objective-driven visualization optimization.
    \item A Neuro-Symbolic Intent Formulation Agent: To the best of our knowledge, we introduce the first neuro-symbolic design for agentic volume visualization. The agent combines MLLM-based semantic interpretation with symbolic reasoning over established clinical knowledge to formulate explicit visualization specifications. Compared with MLLM-only reasoning, the agent provides a more complete set of clinically grounded ROIs in the resulting specifications.
    \item A Multi-Model ROI Identification Agent: We introduce an agent that directly identifies semantically specified ROIs in the original volume by leveraging complementary large-scale pretrained segmentation models. Compared with indirect MLLM-based inference from multi-view renderings, the agent enables more reliable ROI identification while supporting on-demand identification.
    \item An Objective-Driven Visualization Optimization Agent: We introduce an agent that uses a volume-based visibility objective to explicitly evaluate ROI visibility and occlusion in the original volume. Compared with implicit MLLM-based visibility assessment of renderings, the agent enables more reliable ROI-guided volume visualization while improving computational efficiency.
\end{itemize}

\begin{figure*}[ht!]
   \includegraphics[width=\textwidth]{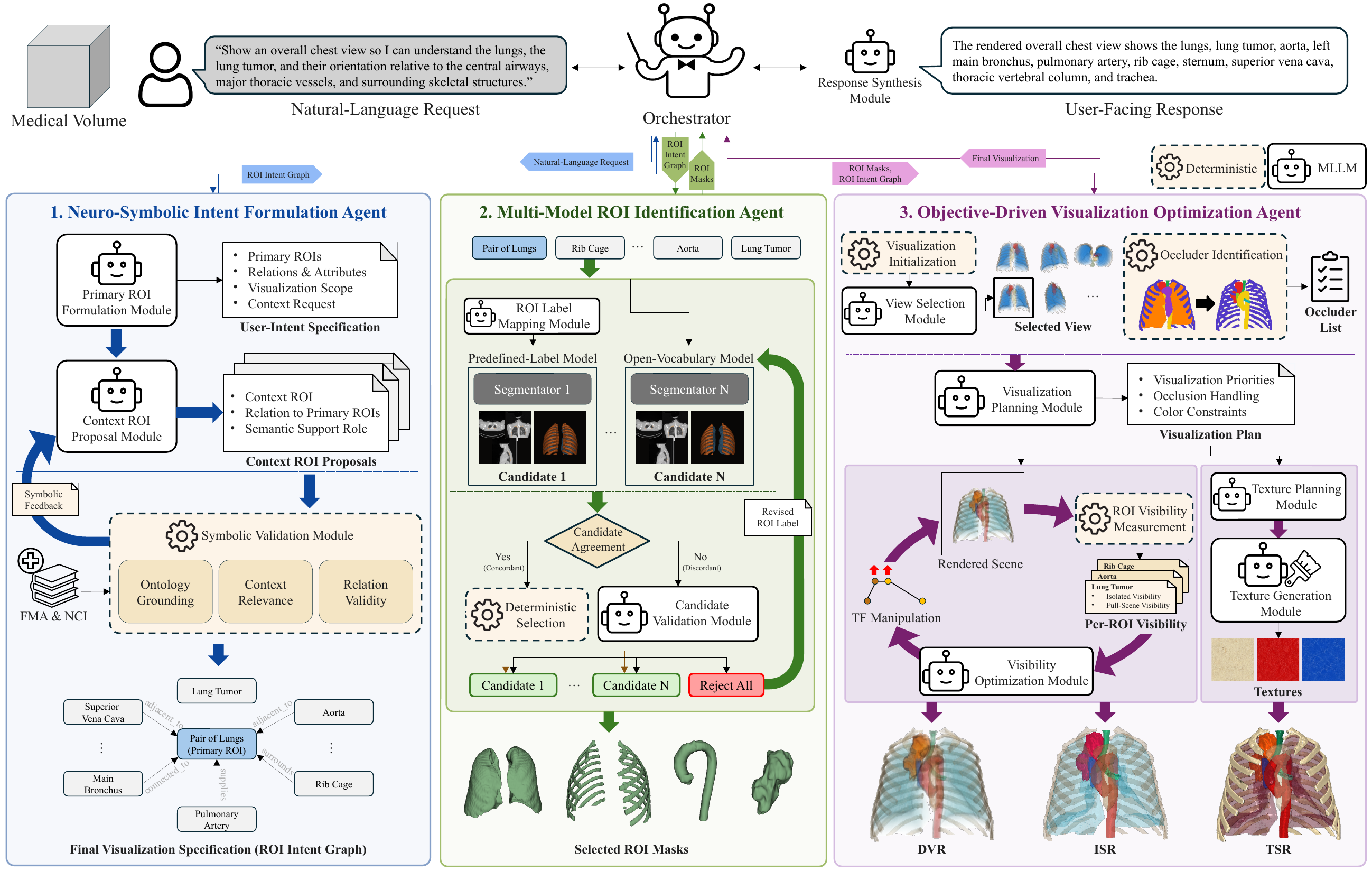}
  \caption{Overview of our MedVA system.}
  \label{fig:fig2}
\end{figure*}

\section{Related works}

\subsection{Medical Volume Visualization}
Medical volume visualization has long incorporated domain knowledge to determine clinically relevant visual content. Schubert et al.\cite{Schubert1993SpatialKnowledgeRepresentation} and Pommert et al.\cite{Pommert1994SymbolicModelingHuman} linked symbolic anatomical knowledge to segmented volumes, Wong et al.\cite{Wong1999SemiAutomaticSceneGeneration} used anatomical ontology relations to construct 3D scenes with related anatomy, and Ezquerra et al.\cite{Ezquerra1999InteractiveKnowledgeGuided} used task-specific clinical rules to interpret cardiac SPECT findings and guide their presentation. These studies showed that explicit medical knowledge can guide content selection and presentation. Once targets are specified, modern segmentation and visualization methods provide strong capabilities to realize the intended visualization. TotalSegmentator\cite{Wasserthal2023TotalSegmentatorRobust} and VoxTell\cite{Rokuss2026VoxTellFreeText} identify predefined or flexibly described ROIs in medical volumes, while importance-, visibility-, and context-aware methods \cite{Viola2004ImportanceDrivenVolume,Correa2011VisibilityHistograms,Bruckner2006IllustrativeContextPreserving} emphasize selected ROIs and preserve relevant context. Thus, prior work provides strong capabilities to select, identify, and present clinically relevant ROIs, but generally starts from explicitly specified target concepts or task-specific clinical settings. The formulation of visualization goals from potentially underspecified clinical requests has received comparatively limited attention.

\newcolumntype{Y}{>{\centering\arraybackslash}X}

\begin{table}[ht]
    \centering
    \caption{Comparison of representative agentic visualization systems.}
    \label{tab:tab1}

    \fontsize{5.8}{6.4}\selectfont
    \setlength{\tabcolsep}{0.55pt}
    \renewcommand{\arraystretch}{1.12}

    \renewcommand{\tabularxcolumn}[1]{m{#1}}

    \begin{tabularx}{\columnwidth}{
        @{}
        >{\centering\arraybackslash}m{0.16\columnwidth}
        Y Y Y Y Y Y
        @{}
    }
        \toprule

        \textbf{Method}
        &
        \textbf{Natural-Language Interaction}
        &
        \textbf{Semantic Target Mapping}
        &
        \textbf{External Domain-Knowledge Integration}
        &
        \textbf{Neuro-Symbolic Validation}
        &
        \textbf{Direct ROI Identification in the Input Volume}
        &
        \textbf{Explicit Volume-based Measurement}
        \\

        \midrule

        ChatVis \cite{Mallick2024ChatVisAutomatingScientific}
        & $\checkmark$
        & $\times$
        & $\times$
        & $\times$
        & $\times$
        & $\times$
        \\

        AVA \cite{Liu2024AVATowardsAutonomous}
        & $\checkmark$
        & $\times$
        & $\times$
        & $\times$
        & $\times$
        & $\times$
        \\

        IntuiTF \cite{Wang2025IntuiTFMLLMGuided}
        & $\checkmark$
        & $\times$
        & $\times$
        & $\times$
        & $\times$
        & $\times$
        \\

        NLI4VolVis \cite{Ai2026NLI4VolVisNaturalLanguage}
        & $\checkmark$
        & $\checkmark$
        & $\times$
        & $\times$
        & $\times$
        & $\times$
        \\

        VizGenie \cite{Biswas2026VizGenieToward}
        & $\checkmark$
        & $\checkmark$
        & $\checkmark$
        & $\times$
        & $\times$
        & $\times$
        \\

        SASAV \cite{Sun2026SASAVSelfDirected}
        & $\times$
        & $\times$
        & $\checkmark$
        & $\times$
        & $\times$
        & $\times$
        \\

        HiLSVA \cite{Ai2026HiLSVADesignEvaluation}
        & $\checkmark$
        & $\times$
        & $\checkmark$
        & $\times$
        & $\times$
        & $\times$
        \\

        \textbf{MedVA (Ours)}
        & $\checkmark$
        & $\checkmark$
        & $\checkmark$
        & $\checkmark$
        & $\checkmark$
        & $\checkmark$
        \\

        \bottomrule
    \end{tabularx}

    \vspace{2pt}

    \parbox{\columnwidth}{%
        \fontsize{5.6}{6.2}\selectfont
        \textit{Note:}
Semantic target mapping links a user-referred target to an explicit semantic component or feature representation. Direct ROI identification locates the ROI in the input volume without a prepared component or predefined segmentation label. Explicit rendering-state measurement denotes renderer-computed ROI visibility or occlusion rather than MLLM inference from rendered images.}
\end{table}

\subsection{Agent System for Volume Visualization}
Agentic visualization research has progressively expanded from natural-language task execution to semantic interaction and autonomous refinement. ChatVis\cite{Mallick2024ChatVisAutomatingScientific} translates natural-language requests into executable visualization scripts and iteratively corrects execution errors, while AVA\cite{Liu2024AVATowardsAutonomous} and IntuiTF\cite{Wang2025IntuiTFMLLMGuided} extend this paradigm as MLLMs assess rendered outputs and guide visualization parameter optimization. This visual feedback enables autonomous refinement, but the assessment of the rendering state remains implicit in the MLLM’s interpretation of rendered images. More recent systems broaden the scope of agentic workflows beyond individual visualization operations. NLI4VolVis\cite{Ai2026NLI4VolVisNaturalLanguage} supports open-vocabulary interaction with editable volumetric scenes, and VizGenie\cite{Biswas2026VizGenieToward} introduces domain-aware feature exploration through vision models and retrieval. However, in these systems, user-referred targets are mapped to prepared semantic components or feature representations rather than identified directly as ROIs in the input volume. SASAV\cite{Sun2026SASAVSelfDirected} and HiLSVA\cite{Ai2026HiLSVADesignEvaluation} further extend agentic workflows through knowledge retrieval, autonomous planning, and broader workflow coordination. Although such knowledge can inform agent reasoning, these systems do not explicitly validate the agent’s interpretation of the visualization intent against structured domain knowledge. \Cref{tab:tab1} summarizes these distinctions among representative agentic visualization systems.

\subsection{Neuro-Symbolic in Agent Systems}
Neuro-symbolic reasoning combines the semantic flexibility of neural models with explicit rules, structured knowledge, or formal reasoning mechanisms. In agent systems, neural models often interpret natural language or propose candidate decisions, while symbolic components verify constraints or perform deterministic reasoning. LLM Modulo \cite{Kambhampati2024LLMsCantPlan} evaluates LLM-generated plans with external critics, Logic-LM \cite{Pan2023LogicLMEmpowering} delegates formal inference to symbolic solvers, and Think-on-Graph \cite{Sun2024ThinkOnGraphDeep} constrains LLM reasoning through knowledge graph traversal. Medical applications follow a similar pattern. MedIKAL \cite{Jia2025medIKALIntegratingKnowledge} incorporates a clinical knowledge graph into LLM-based diagnostic reasoning, and Prenosil et al. \cite{Prenosil2025NeuroSymbolicAIAuditable} use a rule-based expert system to verify information extracted from PET/CT reports. Visualization-related systems have also incorporated structured knowledge for generation or retrieval, as in DIVE \cite{Stutz2025DIVENeuroSymbolic} and NL-2-SPARQL \cite{Codiglione2025NL2SPARQLOntologyBased}. Across these studies, symbolic components primarily support plan verification, logical inference, knowledge retrieval, or output validation. However, neuro-symbolic reasoning remains largely unexplored in agentic volume visualization as a means to validate MLLM interpretations of clinically relevant ROIs and their relationships.

\section{Method}
\subsection{Overview}
\Cref{fig:fig2} presents an overview of MedVA, a multi-agent system that transforms a medical volume and a natural-language request into a volume visualization and a corresponding response. The orchestrator coordinates three agents for clinically grounded intent formulation, direct ROI identification, and objective-driven visualization optimization. The neuro-symbolic intent formulation agent uses MLLM-based modules to interpret the request, specify primary ROIs, and propose context ROIs. An ontology-based symbolic validation module verifies and refines the ROIs and their relationships based on anatomical and pathological knowledge. The agent represents the validated intent as an ROI intent graph. The multi-model ROI identification agent uses multiple pretrained segmentation models to generate candidate masks for the ROIs in the graph. An MLLM-based ROI label mapping module supports predefined-label segmentation models, and deterministic candidate selection returns a mask when the candidate masks satisfy the predefined selection criteria. An MLLM-based candidate validation module evaluates the remaining cases. The objective-driven visualization optimization agent uses MLLM-based modules for view selection and visualization planning and uses explicit occlusion and visibility measurements to evaluate the visualization. For DVR and ISR, an MLLM-based visibility optimization module uses these measurements to guide ROI opacity adjustment. For textured surface rendering (TSR), an MLLM-based texture planning module defines ROI-specific texture specifications, and an image-based texture generation module generates the corresponding textures. Finally, the MLLM-based response synthesis module uses the ROI intent graph and the final visualization to generate a response about the displayed ROIs and their relationships.

\subsection{Neuro-Symbolic Intent Formulation Agent}
The neuro-symbolic intent formulation agent translates a natural-language request into a visualization specification represented as an ROI intent graph. Specifically, the agent formulates the intent expressed in the request and determines whether additional context ROIs are needed to support the visualization goal. The agent consists of two MLLM-based modules and an ontology-based symbolic validation module. The primary ROI formulation module extracts anatomical structures, pathological findings (e.g., tumors and lesions), relations, and attributes (e.g., laterality and multiplicity) from the request. The module designates explicitly requested anatomical structures and pathological findings as primary ROIs and organizes them with the extracted relations and attributes into a user-intent specification. The specification also captures the visualization scope at a global, intermediate, or detailed level based on the anatomical granularity of the request. If the request requires additional context ROIs, the user-intent specification records a context request and the associated primary ROIs. The context ROI proposal module uses this specification to propose context ROIs that provide the required contextual information. Each proposed context ROI specifies its relation to the associated primary ROIs and its semantic support role.

The symbolic validation module verifies the user-intent specification and evaluates the proposed context ROIs against explicit medical knowledge. It uses the Foundational Model of Anatomy (FMA) \cite{Rosse2003ReferenceOntologyBiomedical} as the anatomical knowledge source and the National Cancer Institute Thesaurus (NCI) \cite{Sioutos2007NCIThesaurusSemantic} as the pathological knowledge source. The module checks ontology grounding, context relevance, and relation validity against FMA and NCI. For primary ROIs, it preserves the intent expressed in the request and resolves broad or composite concepts into more specific structures when necessary. For context ROIs, it accepts proposals that satisfy the validation criteria and returns failed proposals to the context ROI proposal module for revision with symbolic feedback. The feedback explains why the proposals failed validation and may provide ontology-grounded alternatives. This neuro-symbolic refinement loop permits up to three revisions. The agent then constructs the ROI intent graph from the validated user-intent specification and context ROIs. Primary and context ROIs form the graph nodes, and validated relations form the edges. The nodes retain relevant attributes, and context ROI nodes also retain their semantic support roles. The ROI intent graph serves as the final visualization specification.

\subsection{Multi-Model ROI Identification Agent}
The multi-model ROI identification agent determines a segmentation mask for each ROI node in the ROI intent graph from the input medical volume. The agent uses multiple segmentation models to generate candidate masks for each ROI. These models differ in their input requirements, so the agent adapts each ROI input to the corresponding model. Open-vocabulary models directly receive the ROI label as input. For predefined-label models, an MLLM-based ROI label mapping module compares the ROI label with the model's predefined label set and selects a semantically compatible label. If the module cannot find a compatible label, it returns no mapping, and no candidate mask is generated from that model for the ROI.

The candidate masks generated for an ROI can differ across segmentation models. The agent therefore evaluates agreement among the candidate masks before selecting a final mask. It computes Dice scores for all candidate pairs and identifies the pair with the highest score. A high Dice score indicates that two segmentation models produce similar masks for the same ROI. This cross-model comparison reduces reliance on a single model prediction. If the highest score satisfies candidate-agreement threshold, the agent compares the two candidates based on image-boundary consistency. The agent computes a gradient map from the input medical volume and measures the average gradient magnitude along the boundary of each candidate mask. Because the two candidates already have substantial spatial overlap, this measure provides an additional image-based criterion for comparing their differing boundaries. The agent selects the candidate with the higher boundary-consistency score. Otherwise, the full candidate set is forwarded to the MLLM-based candidate validation module. For each candidate, the module prepares three orthogonal slice overlays and a 3D mesh rendering. All candidate visualizations use the same viewing configuration. The module evaluates whether each candidate corresponds to the requested ROI and exhibits a plausible spatial extent and shape in the input volume. It selects the candidate that best matches the requested ROI or rejects all candidates when none satisfies the validation criteria. For an ROI whose candidate set is rejected, the validation module explains the failure and proposes a revised ROI label based on the ROI attributes and relations in the ROI intent graph. The open-vocabulary models regenerate candidates from the revised label, and the agent reevaluates them together with the existing candidates from predefined-label models. This refinement continues until a candidate is selected or the maximum number of attempts is reached

\subsection{Objective-Driven Visualization Optimization Agent}
The objective-driven visualization optimization agent uses the medical volume, ROI intent graph, and ROI masks to produce an ROI-guided visualization according to the specified ROI priorities. The agent first initializes the visualization by assigning red to primary ROIs and distinct colors from a predefined palette to context ROIs. The MLLM-based view selection module then evaluates ten predefined anatomical views based on how well each view exposes the primary ROIs and satisfies the orientation requirements specified in the ROI intent graph. The selected view is shared across DVR, ISR, and TSR. From the selected view, the agent determines how much of each ROI is occluded in the full-scene rendering. If the occlusion exceeds a predefined threshold, the agent evaluates the other ROIs individually and designates each ROI that contributes to the occlusion as an occluder. The MLLM-based visualization planning module uses the ROI intent graph and the occluder list to construct a visualization plan that includes information such as visualization priorities, occlusion handling, and color constraints. The visualization plan then guides DVR, ISR, and TSR.

For DVR and ISR, ROI opacity is refined through volume-based VPA loop. The agent derives initial opacity values and visibility requirements for each ROI from the visualization plan. For each ROI, the agent measures visibility in a full-scene rendering and an isolated rendering. The visibility when the ROI is rendered alone indicates whether its opacity is sufficient. The reduction from this value to the full-scene visibility indicates visibility loss caused by occlusion from other ROIs. ROI visibility is calculated as the mean front-to-back opacity contribution across all viewing rays that intersect the ROI:

\begin{equation}
V_r =
\frac{1}{|P_r|}
\sum_{p \in P_r}
\sum_{k \in I_r(p)}
T_k \alpha_k,
\qquad
T_k = \prod_{j<k}(1-\alpha_j).
\label{eq:roi_visibility}
\end{equation}
where $P_r$ denotes the set of viewing rays that intersect ROI $r$, $I_r(p)$ is the set of sample indices belonging to ROI $r$ along the ray  $p$, $\alpha_k$ is the opacity of sample $k$, and $T_k$ is the accumulated transmittance before that sample.

The MLLM-based visibility optimization module receives the full-scene and isolated visibility measurements together with the visualization plan and visibility requirements. At each iteration, the module identifies an ROI that does not satisfy its visibility requirement and determines whether the deficit results from insufficient ROI opacity or occlusion. It then proposes an opacity adjustment based on the identified cause and the visualization plan. The agent applies the adjustment and recomputes the visibility measurements. The adjustment is retained only if it improves the target ROI's visibility without causing other ROIs to violate their visibility requirements. The optimization continues until all visibility requirements are satisfied or the maximum number of iterations is reached.

For TSR, the MLLM-based texture planning module converts the visualization plan into ROI-specific texture specifications. Each specification defines the texture pattern and surface style of the corresponding ROI based on its role and relationships. The specification also defines reduced opacity when the ROI is designated as an occluder. The texture generation module generates a texture map from each specification and applies it to the corresponding ROI surface.

\subsection{Interactive Workflow Orchestration}
The orchestrator coordinates agent execution for each user interaction and manages the ROI intent graph, identified ROI masks, visualization plan, and current visualization. For each follow-up request, it determines which agents need to be invoked and which existing results can be reused. When a user requests an ROI addition, the orchestrator invokes the intent formulation and ROI identification agents only for the new ROI and updates the current visualization with the resulting mask. A request to change color or opacity updates only the corresponding visualization parameter. A question about the current visualization invokes only the MLLM-based response synthesis module without changing the visualization. The response synthesis module uses the ROI intent graph and current visualization to answer questions about the displayed ROIs, their appearance, and their anatomical relationships. Users can also change the rendering mode, viewpoint, clipping, color, and opacity through interface controls. These changes directly update the current visualization without invoking the visualization optimization agent.

\subsection{System Implementation}
\textbf{MLLM Prompting and structured outputs} Each MLLM-based module uses role-specific instructions that define its task, allowed decisions, and expected output format. The primary ROI formulation, context ROI proposal, and ROI label mapping include compact examples that support in-context learning for their respective tasks. All MLLM outputs follow predefined JSON schemas and are validated before further processing. When the required information cannot be determined from the provided inputs, the prompts instruct the module to return an explicit failure status rather than generate an unsupported output.

\textbf{Interactive interface} MedVA provides an interactive interface that integrates 3D visualization, direct controls, natural-language interaction, and the agent execution log. MedVA also provides a 1D transfer function that allows users to overlay selected intensity ranges from the original volume on the ROI rendering. This supports the joint visualization of segmented ROIs and information not represented by the available ROI masks, such as calcifications, as shown in \Cref{fig:fig1}. The interface is demonstrated in the supplementary video.

\begin{table}[!t]
  \caption{ROI specifications with and without the symbolic validation module for natural-language requests. Primary ROIs are bold; the right column lists only additions or refinements, with unlisted ROIs retained.}
  \label{tab:tab2}

  \centering
  \fontsize{7.5}{8.2}\selectfont
  \setlength{\tabcolsep}{3pt}
  \renewcommand{\arraystretch}{1.00}

  \begin{tabularx}{\columnwidth}{
    @{}
    >{\raggedright\arraybackslash}X
    >{\raggedright\arraybackslash}X
    @{}
  }

    \toprule
    \multicolumn{1}{c}{\textbf{Without Neuro-Symbolic}} &
    \multicolumn{1}{c}{\textbf{With Neuro-Symbolic}} \\
    \midrule

    \rowcolor{black!7}
    \multicolumn{2}{@{}p{\columnwidth}@{}}{%
      Prompt: Show the pancreas and the major vessels running around it.
    } \\[1pt]

    \textbf{pancreas}, splenic artery, splenic vein,
    superior mesenteric artery, superior mesenteric vein, portal vein
    &
    \textit{(No change)}
    \\[2pt]

    \midrule

    \rowcolor{black!7}
    \multicolumn{2}{@{}p{\columnwidth}@{}}{%
      Prompt: Show the cardiac chambers.
    } \\[1pt]

    \textbf{cardiac chambers}
    &
    \textit{(Refinement)} \newline
    \sout{\textbf{cardiac chambers}} $\rightarrow$ \newline
    \textbf{left atrium},
    \textbf{left ventricle},
    \textbf{right atrium},
    \textbf{right ventricle}
    \\[2pt]

    \midrule

    \rowcolor{black!7}
    \multicolumn{2}{@{}p{\columnwidth}@{}}{%
      Prompt: Show the ventricular system with brain landmarks.
    } \\[1pt]

    \textbf{ventricular system}, corpus callosum,
    thalamus, brainstem, cerebellum
    &
    \textit{(Refinement)} \newline
    \sout{\textbf{ventricular system}} $\rightarrow$ \newline
    \textbf{lateral ventricles},
    \textbf{third ventricle},
    \textbf{fourth ventricle}
    \\[2pt]

    \midrule

    \rowcolor{black!7}
    \multicolumn{2}{@{}p{\columnwidth}@{}}{%
      Prompt: Show the heart with structures needed for cardiac orientation.
    } \\[1pt]

    \textbf{heart}, ascending aorta, pulmonary trunk,
    superior vena cava, inferior vena cava, pericardium
    &
    \textit{(Addition)} \newline
    left coronary artery,
    right coronary artery
    \\

    \bottomrule

  \end{tabularx}
\end{table}

\section{Evaluation Setup}
We evaluated MedVA through agent-level experiments and integrated case studies. The agent-level experiments assessed the three agents separately. For neuro-symbolic intent formulation, we compared ROI specifications generated with and without the symbolic validation module across four natural-language visualization requests. For ROI identification, we instantiated the multi-model agent with TotalSegmentator \cite{Wasserthal2023TotalSegmentatorRobust} and VoxTell \cite{Rokuss2026VoxTellFreeText}, which represent predefined-label and open-vocabulary segmentation, respectively. We selected this pair to cover the two complementary segmentation paradigms supported by the agent. We evaluated their ROI-wise segmentation performance and examined how the agent combined their outputs in chest and liver CT volumes. For objective-driven visualization optimization, we examined primary ROI-guided optimization under DVR and ISR and compared the rendering-based and volume-based VPA loops across pelvic, liver, and chest CT volumes \cite{Ferrara2026SharingWholeTotalBody, IRCAD3DIRCADb01, Sang2026BenchmarkSegmentationPelvic} in terms of optimization time, token consumption, and iteration count. The case studies examined varied user intents, multi-step user interactions, consistency across patients and datasets, and end-to-end execution time.

For the VPA comparison, we compared a rendering-based VPA loop with our volume-based VPA loop. The two conditions differed only in the information provided to the MLLM. The rendering-based loop received intermediate renderings, whereas the volume-based loop received quantitative ROI visibility measures. For each condition, the VPA optimization was limited to six iterations, and the entire optimization was repeated five times.

Across experiments, the candidate-agreement threshold was set to a Dice score of 0.75, and the occluder-overlap threshold was set to 0.60. In TSR, non-occluding ROIs used an opacity of 1.0 and designated occluders used an opacity of 0.5. The system used GPT 5.6 Terra for the language and multimodal modules, GPT Image 2 for texture generation, VoxTell 0.1.1 and TotalSegmentator 2.13.0 for ROI identification, and FMA 5.1.0 and NCI 26.07d for symbolic grounding. Experiments ran on Ubuntu 24.04 with an Intel Core i7-14700K, 64 GB RAM, and an NVIDIA RTX 4090 with 24 GB VRAM.

\begin{table*}[t]
    \caption{ROI-wise performance of two large-scale pretrained medical segmentation models across thirteen representative ROIs. Values report the mean and standard deviation of Dice in percent; bold indicates the higher mean. Bilateral structures are summarized by equally weighted pooling of the reported left- and right-side distributions. Iliac A. denotes the iliac arteries, and Iliac V. denotes the iliac veins.}
    \label{tab:tab3}
    \centering
    \fontsize{7.5}{8.5}\selectfont
    \setlength{\tabcolsep}{1.25pt}
    \renewcommand{\arraystretch}{1.25}

    \begin{tabular*}{\textwidth}{
        @{\extracolsep{\fill}}
        l
        *{13}{c}
        @{}
    }
        \toprule
        \textbf{Method}
        & \textbf{Lungs}
        & \textbf{Kidneys}
        & \textbf{Liver}
        & \textbf{Pancreas}
        & \textbf{Spleen}
        & \textbf{Stomach}
        & \textbf{Bladder}
        & \textbf{Aorta}
        & \textbf{IVC}
        & \textbf{Iliac A.}
        & \textbf{Iliac V.}
        & \textbf{Bone}
        & \textbf{Tumor} \\
        \midrule

        TotalSeg.
        & $95.6 \pm 2.9$
        & $93.0 \pm 10.2$
        & $97.2 \pm 0.7$
        & $86.3 \pm 10.4$
        & $95.8 \pm 1.2$
        & $\mathbf{94.1 \pm 2.3}$
        & $82.0 \pm 18.0$
        & $87.6 \pm 7.9$
        & $78.1 \pm 10.7$
        & $\mathbf{91.4 \pm 3.8}$
        & $\mathbf{93.4 \pm 3.0}$
        & $80.6 \pm 5.8$
        & $58.0 \pm 27.0$ \\

        VoxTell
        & $\mathbf{96.0 \pm 2.4}$
        & $\mathbf{94.0 \pm 10.8}$
        & $\mathbf{97.5 \pm 0.7}$
        & $\mathbf{88.1 \pm 6.5}$
        & $\mathbf{96.6 \pm 1.3}$
        & $93.3 \pm 4.6$
        & $\mathbf{85.8 \pm 15.3}$
        & $\mathbf{89.5 \pm 7.4}$
        & $\mathbf{80.1 \pm 11.0}$
        & $56.6 \pm 14.1$
        & $84.3 \pm 4.4$
        & $\mathbf{85.8 \pm 5.7}$
        & $\mathbf{87.5 \pm 4.2}$ \\
        \bottomrule
    \end{tabular*}
\end{table*}

\section{Agents Evaluation}

\subsection{Neuro-Symbolic Intent Formulation Agent}
\Cref{tab:tab2} presents the ROI specifications from the intent formulation agent across four volumes with diverse ROIs and user requests at varying levels of specificity and illustrates the contribution of the symbolic validation module. Overall, the MLLM-only baseline agent produced ROI specifications that were broadly aligned with the user intents, while the symbolic validation module selectively refined them when greater clinical specificity or completeness was needed. For the sufficiently specific pancreas request, the baseline agent could already produce a complete ROI specification. In the remaining cases, the symbolic validation module refined the ROI specification either by decomposing broad concepts into more clinically specific substructures or by adding clinically relevant context ROIs. For example, the general term “cardiac chambers” was decomposed into the left atrium, left ventricle, right atrium, and right ventricle. For the cardiac orientation request, the left and right coronary arteries were additionally introduced as context ROIs, while the previously specified ROIs were retained.

\begin{figure}[t]
  \centering
  \includegraphics[width=\linewidth]{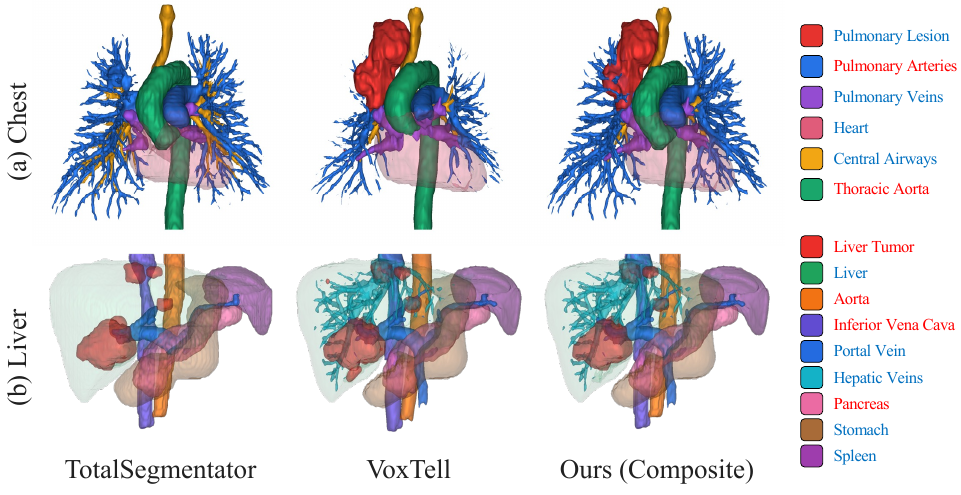}
  \caption{ROI identification results for (a) chest and (b) liver volumes\cite{Ferrara2026SharingWholeTotalBody,IRCAD3DIRCADb01}. In the composite results, red labels denote ROIs selected from TotalSegmentator and blue labels denote those selected from VoxTell.}
  \label{fig:fig3}
\end{figure}

\subsection{Multi-Model ROI Identification Agent}
\Cref{tab:tab3} presents the ROI-wise segmentation performance of TotalSegmentator and VoxTell, which are employed by the multi-model ROI identification agent. These two large-scale pretrained medical segmentation models were evaluated across thirteen representative ROIs encompassing organs, major vessels, skeletal structures, and lesions. Overall, both models achieved strong and comparable segmentation performance, with mean Dice scores of 87.16\% and 87.32\% for TotalSegmentator and VoxTell, respectively. Neither model consistently outperformed the other across all ROIs. Instead, each showed strengths for different ROIs, which indicates that their segmentation capabilities are complementary. In particular, substantial performance differences were observed for the iliac arteries, iliac veins, and tumors. For these ROIs, the better-performing model achieved markedly higher Dice scores, such as 91.4\% for the iliac arteries with TotalSegmentator and 87.5\% for tumors with VoxTell.

\Cref{fig:fig3} illustrates the results of the ROI identification agent across two medical volumes containing diverse anatomical and pathological ROIs. Overall, both large-scale pretrained medical segmentation models identified most requested ROIs, but neither model alone provided the complete ROI set. Their outputs were complementary, and our agent evaluated the candidate segmentations using the deterministic selection and MLLM-based validation modules and combined the appropriate model-specific results to form the final ROI sets. For the chest volume (\cref{fig:fig3}(a)), most requested ROIs were identified by both models, while the pulmonary lesion was detected only by VoxTell. Among the ROIs identified by both models, the agent selected the TotalSegmentator results for the pulmonary arteries and thoracic aorta because they provided more complete segmentations. For the central airway, the VoxTell result was selected because the TotalSegmentator result extended excessively into the peripheral bronchi. A similar complementary pattern was observed for the liver volume (\cref{fig:fig3}(b)). The agent selected four ROIs from TotalSegmentator and the remaining five from VoxTell, including the hepatic veins identified only by the latter.

\begin{table}[!ht]
  \caption{Optimization efficiency comparison between the rendering-based VPA loop and our volume-based VPA loop. Time and tokens are reported as mean $\pm$ SD; Iters denotes mean iteration count.}
  \label{tab:tab4}
  \centering
  \fontsize{7.0}{7.6}\selectfont
  \setlength{\tabcolsep}{2.5pt}
  \renewcommand{\arraystretch}{1.00}

  \begin{tabular*}{\columnwidth}{
    @{\extracolsep{\fill}}
    cccccc
    @{}
  }    
    \toprule
    \textbf{Volume} &
    \textbf{Mode} &
    \textbf{VPA loop} &
    \textbf{Iters} &
    \textbf{Seconds} &
    \textbf{Tokens (k)} \\
    \midrule

    \multirow[c]{4}{*}{Pelvis}
    & \multirow[c]{2}{*}{DVR}
    & Rendering-based
    & 6.0
    & $28.67 \pm 3.28$
    & $11.984 \pm 0.094$ \\

    &
    & Volume-based
    & 1.0
    & $9.19 \pm 0.19$
    & $1.590 \pm 0.008$ \\

    \cmidrule(lr){2-6}

    & \multirow[c]{2}{*}{ISR}
    & Rendering-based
    & 2.0
    & $11.21 \pm 14.09$
    & $3.890 \pm 3.474$ \\

    &
    & Volume-based
    & 1.0
    & $9.67 \pm 0.60$
    & $1.615 \pm 0.003$ \\

    \midrule

    \multirow[c]{4}{*}{Liver}
    & \multirow[c]{2}{*}{DVR}
    & Rendering-based
    & 6.0
    & $31.50 \pm 1.64$
    & $16.199 \pm 0.196$ \\

    &
    & Volume-based
    & 1.0
    & $7.48 \pm 0.87$
    & $2.263 \pm 0.003$ \\

    \cmidrule(lr){2-6}

    & \multirow[c]{2}{*}{ISR}
    & Rendering-based
    & 5.0
    & $27.00 \pm 5.09$
    & $12.601 \pm 0.172$ \\

    &
    & Volume-based
    & 4.0
    & $21.24 \pm 2.41$
    & $10.151 \pm 0.025$ \\

    \midrule

    \multirow[c]{4}{*}{Chest}
    & \multirow[c]{2}{*}{DVR}
    & Rendering-based
    & 5.4
    & $30.39 \pm 10.62$
    & $12.648 \pm 3.257$ \\

    &
    & Volume-based
    & 3.0
    & $16.54 \pm 2.37$
    & $6.529 \pm 0.146$ \\

    \cmidrule(lr){2-6}

    & \multirow[c]{2}{*}{ISR}
    & Rendering-based
    & 2.2
    & $13.25 \pm 7.09$
    & $4.990 \pm 2.538$ \\

    &
    & Volume-based
    & 1.0
    & $7.33 \pm 0.50$
    & $1.893 \pm 0.014$ \\

    \bottomrule
  \end{tabular*}
\end{table}

\begin{figure}[!ht]
  \centering
  \includegraphics[width=\linewidth]{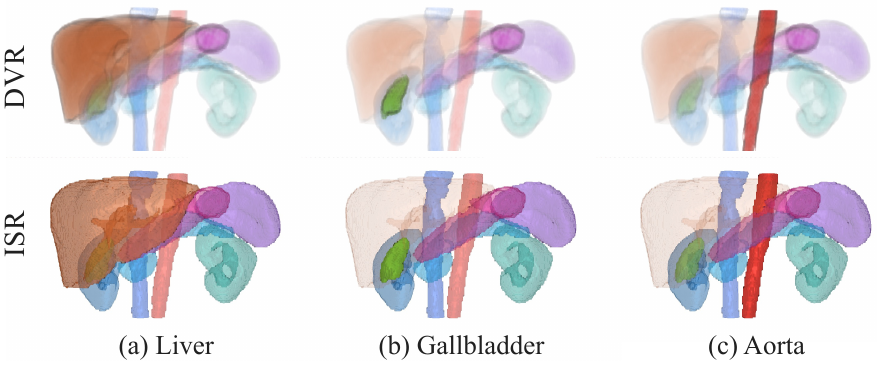}
  \caption{Primary ROI-guided visualization optimization results for the liver, gallbladder, and aorta under DVR and ISR.}
  \label{fig:fig4}
\end{figure}

\subsection{Objective-Driven Visualization Optimization Agent}
\Cref{fig:fig4} presents the primary ROI-guided results of the visualization optimization agent under two different rendering modes, DVR and ISR. Overall, the agent produced optimized visualizations that emphasized the primary ROIs while appropriately adjusting the visibility of the context ROIs according to their spatial and occlusion relationships. For the liver and aorta, only minor adjustments were observed because both primary ROIs were positioned toward the front of the current view and were minimally occluded. In contrast, greater attenuation of the context ROIs was observed for the gallbladder because it was more deeply located and heavily occluded. Both rendering modes showed a consistent ROI-guided optimization pattern, although the changes to context ROIs were less pronounced under ISR due to its iso-surface representation.

\Cref{tab:tab4} compares the optimization efficiency of the rendering-based VPA loop and our volume-based VPA loop across three medical volumes and two rendering modes. The volume-based loop improved optimization efficiency across all three measured metrics. Across the six conditions, the volume-based loop required an average of 11.91 seconds, 4.007k tokens, and 1.83 iterations. These values corresponded to reductions of 49.7\% in optimization time and 61.4\% in token consumption compared with the rendering-based loop. The volume-based loop also substantially reduced the number of iterations. It completed optimization in a single iteration in four conditions, whereas the rendering-based loop frequently approached the maximum of six iterations. 

\begin{figure}[t]
  \centering
  \includegraphics[width=\linewidth]{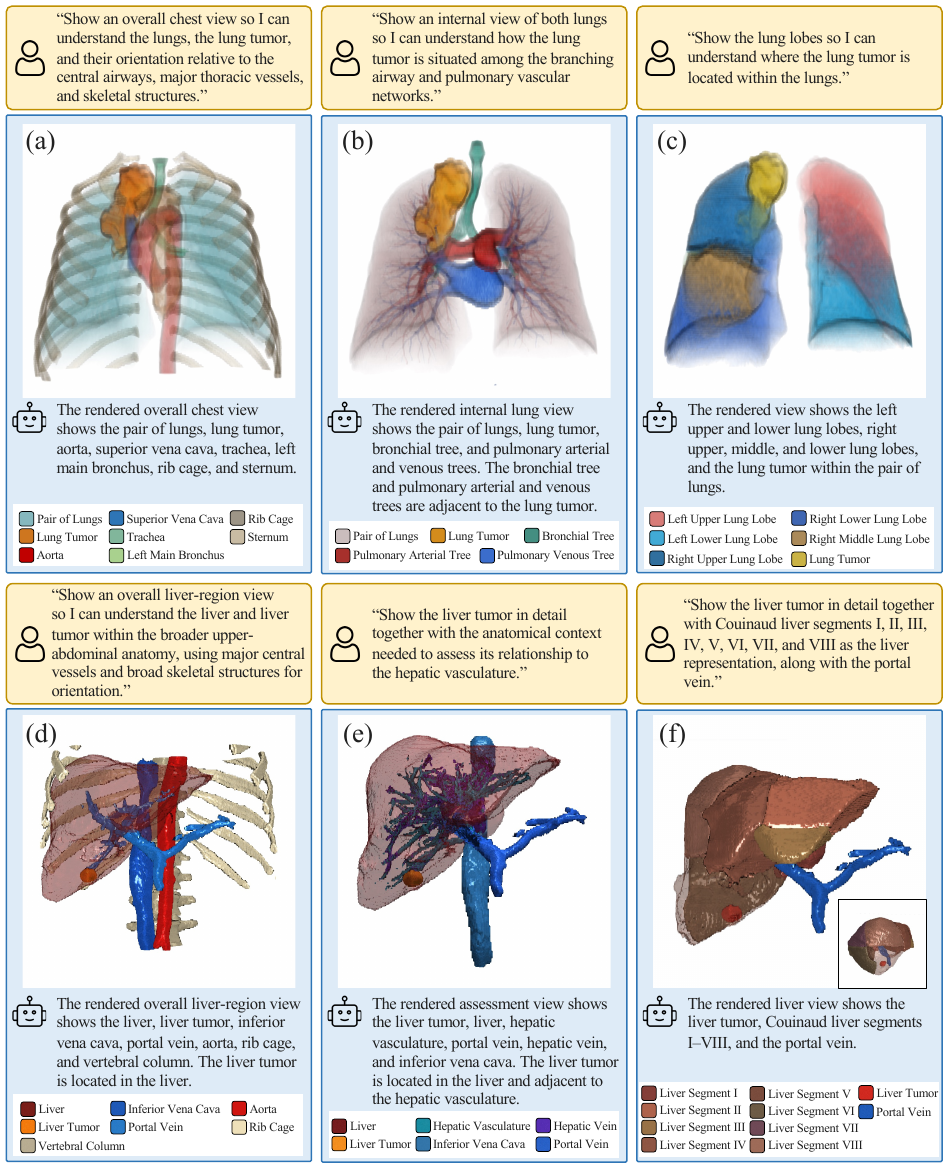}
  \caption{Visualization results for user intents of varying specificity in the chest (a--c) and liver (d--f) volumes\cite{Ferrara2026SharingWholeTotalBody,IRCAD3DIRCADb01}.}
  \label{fig:fig5}
\end{figure}

\begin{figure}[t]
  \centering
  \includegraphics[width=\linewidth]{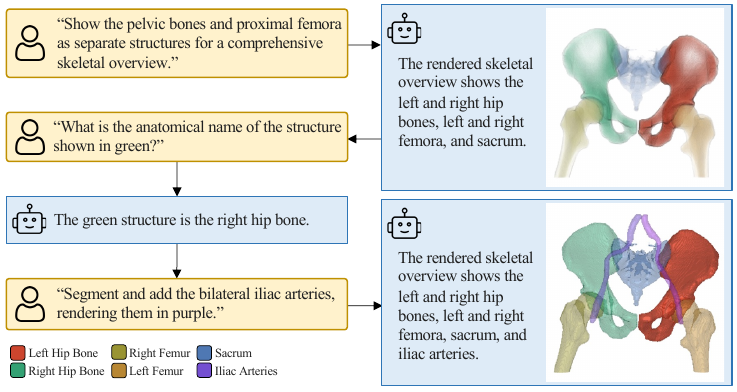}
  \caption{Multi-step interaction in the pelvic volume\cite{Sang2026BenchmarkSegmentationPelvic} combining visualization querying, incremental ROI addition, and rendering mode change.}
  \label{fig:fig6}
\end{figure}

\section{Case Studies}
\subsection{Various User Intents}
\Cref{fig:fig5} presents the visualization results of MedVA for user intents ranging from broad overview requests to focused and clinically specific goals across two medical volumes. MedVA specified and identified the ROIs corresponding to each user intent and visualized them based on visualization priorities and with distinct colors. For a broad request to inspect major thoracic structures in the chest volume, MedVA specified and identified eight ROIs and produced an overview visualization in which all ROIs were readily visible (\cref{fig:fig5}(a)). When the user requested a more focused visualization of the lung tumor together with the surrounding airway and vascular networks, MedVA increased their visual prominence while excluding the irrelevant aorta and rib cage (\cref{fig:fig5}(b)). For a more clinically specific goal describing the tumor location relative to the lung lobes, the resulting visualization emphasized the tumor together with the five lung lobes and clearly depicted its location within the right upper lobe (\cref{fig:fig5}(c)).

The liver volume showed a similar intent-dependent variation. Under the broad overview request, MedVA emphasized the liver and tumor together with major abdominal vessels and skeletal landmarks to provide anatomical context for the tumor (\cref{fig:fig5}(d)). When the request instead focused on the relationship between the tumor and hepatic vasculature, MedVA shifted the visualization emphasis toward these ROIs while excluding the rib cage (\cref{fig:fig5}(e)). For a clinically specific goal describing the tumor location using Couinaud liver segments, MedVA prioritized the tumor and individual liver segments as well as the portal vein (\cref{fig:fig5}(f)).

\subsection{Various User Interactions}
\begin{figure}[t]
  \centering
  \includegraphics[width=\linewidth]{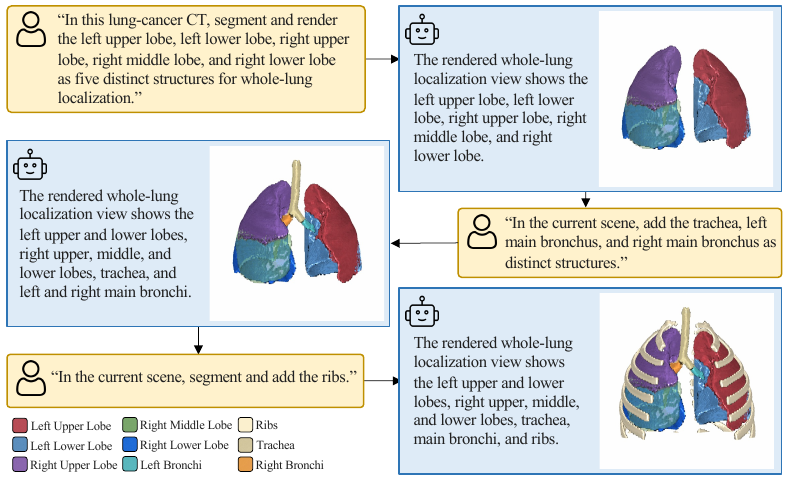}
  \caption{Multi-step interaction in the lung volume\cite{Ferrara2026SharingWholeTotalBody} showing progressive ROI addition.}
  \label{fig:fig7}
\end{figure}

\Cref{fig:fig6,fig:fig7} present multi-step interaction results of MedVA using two medical volumes. For the pelvic volume (\cref{fig:fig6}), MedVA interpreted the initial broad request as five skeletal substructures and produced an overview visualization in which all ROIs were clearly distinguished. The subsequent user interaction was a query on the current visualization, and MedVA correctly recognized the green structure as the right hip bone. In the final request to add bilateral iliac arteries in a user-defined color, MedVA visualized them accordingly while preserving the previous visualization context. The user then changed the rendering mode from DVR to ISR. Unlike the pelvic case, the interaction with the lung volume (\cref{fig:fig7}) began with an explicit request for five individual lung lobes, and MedVA identified and visualized each lobe. In response to subsequent requests, MedVA first added the trachea and main bronchi and then the ribs while preserving the previously visualized ROIs at each interaction step. All visualizations for the chest volume were rendered using TSR.

\subsection{Various Medical Datasets}
\begin{figure}[t]
  \centering
  \includegraphics[width=\linewidth]{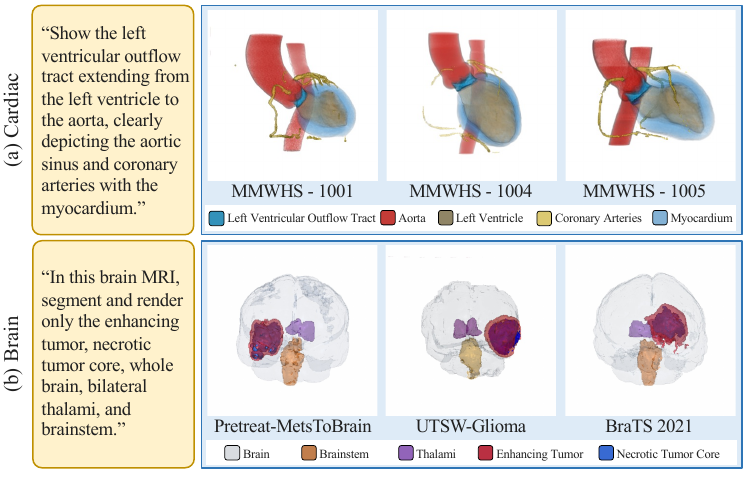}
  \caption{Visualization results for the same user requests across (a) three MM-WHS\cite{Zhuang2019EvaluationAlgorithmsMultiModality} cardiac patients and (b) three brain datasets: Pretreat-MetsToBrain\cite{Ramakrishnan2024LargeOpenAccess}, UTSW-Glioma\cite{Reddy2026UniversityTexasSouthwestern}, and BraTS 2021\cite{Baid2021RSNAASNRMICCAIBraTS}.}
  \label{fig:fig8}
\end{figure}

\Cref{fig:fig8} presents the visualization results of MedVA under inter-patient and inter-dataset variations using the same user requests. For the cardiac volumes from three different patients within the same dataset (\cref{fig:fig8}(a)), MedVA consistently identified all five ROIs and produced comparable visualizations. Similar consistency was observed across the three different brain datasets (\cref{fig:fig8}(b)). Despite substantial variations in ROI location and morphology across the three datasets, MedVA consistently identified all specified ROIs and produced comparable tumor-focused visualizations.

\begin{figure}[t]
  \centering
  \includegraphics[width=\linewidth]{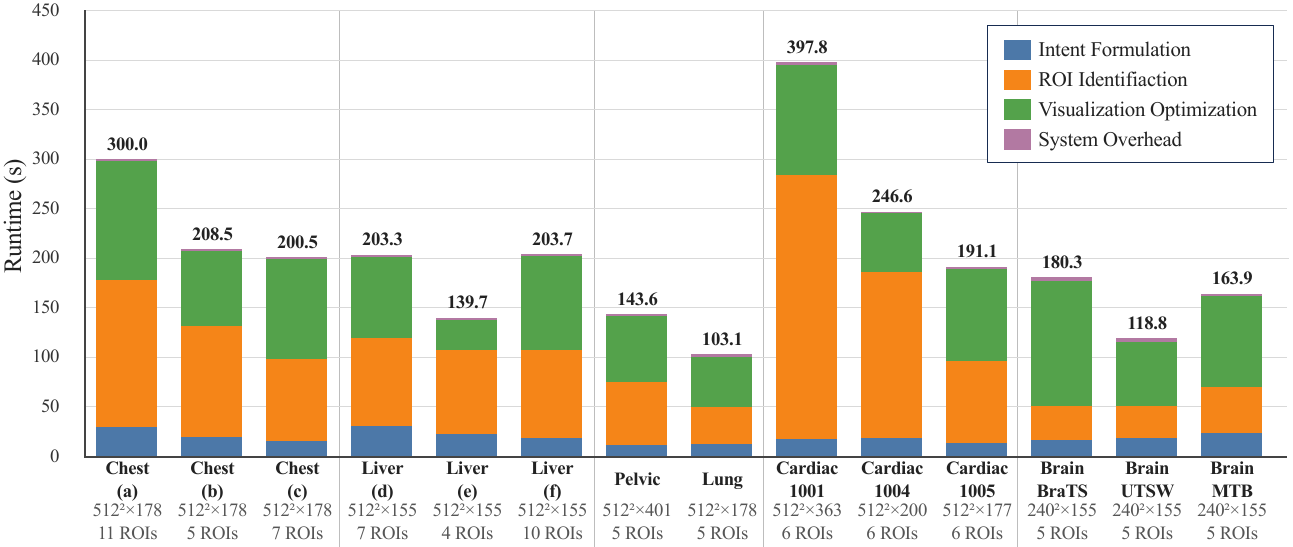}
  \caption{End-to-end runtime (seconds) of MedVA with agent-level breakdown for fourteen cases. Pelvic and lung runtimes are for the first step of the multi-step interactions. Volume size and ROI count are shown below.}
  \label{fig:fig9}
\end{figure}

\subsection{End-to-End Runtime}
\Cref{fig:fig9} presents the end-to-end runtime of MedVA and its agent-level breakdown for the fourteen cases in Sections 6.1-6.3. The end-to-end runtime averaged 200.1 seconds and ranged from 103.1 to 397.8. The variation suggested that runtime depends not only on volume characteristics, such as volume size and identified ROI count, but also on the user request. At the agent level, the ROI identification and visualization optimization agents accounted for most of the total runtime, with an average contribution of 89.4\%.

\section{User Study}
\subsection{Experiment Setup}
We conducted a user study with nine participants. The novice group consisted of seven postgraduate students with less than three years of relevant experience, including four with backgrounds in medical visualization and three in AI and computer vision. The researcher group consisted of two medical researchers with at least five years of relevant experience in medical visualization and clinical implementation. Each participant completed the study independently using the same workstation environment. The study followed the University’s Institutional Review Board protocol, and informed consent was obtained from all participants before the study.

\subsection{Procedure and Tasks}
Participants first completed a background questionnaire and were introduced to MedVA and its overall interaction workflow. They then practiced formulating a visualization intent through natural-language requests, inspecting the resulting visualization, and making follow-up refinements using a separate tutorial volume.

The user study comprised two complementary tasks. Study 1 evaluated how well MedVA visualizations reflected the corresponding user intents using the six cases from Section 6.1. For each case, participants first answered a multiple-choice question that assessed their understanding of the spatial relationship between the primary ROI and relevant context ROIs. They then rated three aspects on five-point Likert scales: alignment between the visualization and user intent (intent alignment), inclusion of the primary and relevant context ROIs to support the visualization goal (ROI inclusion), and ease of interpreting the primary ROI location relative to the context ROIs (spatial interpretation). 

In Study 2, we evaluated whether participants could achieve the desired visualizations using their own freely expressed natural-language requests. Participants were asked to complete the three predefined multi-step interaction scenarios from Section 6.2 and \Cref{fig:fig1}. We recorded task completion for each scenario based on predefined criteria. Participants then rated three aspects on five-point Likert scales: accuracy of the changes produced in response to follow-up requests (request fulfillment), preservation of previous visualization results (visualization preservation), and ease of expressing the desired changes through natural language (expression ease). 

After both studies, participants rated six aspects of their overall user experience (Q1–Q6) on five-point Likert scales. Participants also provided open-ended feedback on useful features, interaction difficulties, and desired improvements. The complete study protocol and questionnaire are provided in the supplementary material. \Cref{tab:tab5} summarizes the task performance and overall user experience results.

\begin{table}[t]
\caption{User study results for (a) task performance and (b) overall user experience. Counts indicate correct responses or completed scenarios; ratings are mean $\pm$ SD ($n=9$).}
\label{tab:tab5}

\centering
\footnotesize
\fontsize{7.0}{7.6}\selectfont
\setlength{\tabcolsep}{2.5pt}
\renewcommand{\arraystretch}{1.00}

\begin{tabularx}{\columnwidth}{
    @{}
    >{\raggedright\arraybackslash}X
    r
    @{\hspace{7pt}}
    >{\raggedright\arraybackslash}X
    r
    @{}
}
\toprule
\multicolumn{4}{c}{\textbf{(a) Task Performance}} \\
\midrule

\multicolumn{2}{c}{\textbf{Study 1}} &
\multicolumn{2}{c}{\textbf{Study 2}} \\
\cmidrule(r){1-2}
\cmidrule(l){3-4}

Response Accuracy
    & 51/54
    & Task Completion
    & 27/27 \\

Intent Alignment
    & $4.87 \pm 0.22$
    & Request Fulfillment
    & $4.81 \pm 0.29$ \\

ROI Inclusion
    & $4.87 \pm 0.18$
    & Visualization Preservation
    & $4.85 \pm 0.34$ \\

Spatial interpretation
    & $5.00 \pm 0.00$
    & Expression Ease
    & $4.89 \pm 0.24$ \\

\bottomrule
\end{tabularx}

\vspace{5pt}

\begin{tabular*}{\columnwidth}{
    @{\extracolsep{\fill}}
    cccccc
    @{}
}
\toprule
\multicolumn{6}{c}{\textbf{(b) Overall User Experience}} \\
\midrule

\textbf{Q1} &
\textbf{Q2} &
\textbf{Q3} &
\textbf{Q4} &
\textbf{Q5} &
\textbf{Q6} \\

$5.00\!\pm\!0.00$ &
$4.78\!\pm\!0.44$ &
$4.89\!\pm\!0.33$ &
$5.00\!\pm\!0.00$ &
$4.89\!\pm\!0.33$ &
$4.67\!\pm\!0.71$ \\

\bottomrule
\end{tabular*}

\vspace{2pt}

\parbox{\columnwidth}{%
    \fontsize{5.8}{6.4}\selectfont
    \textit{Note:}
    Q1, ease of learning;
    Q2, results without manual parameter adjustment;
    Q3, goal expression without explicit ROI naming;
    Q4, intuitive natural-language interaction;
    Q5, support for understanding ROI relationships;
    Q6, intention to use MedVA for similar tasks.
}
\end{table}

\subsection{Task Performance}
Participants answered 51 of 54 multiple-choice questions in Study 1 correctly (94.4\%). All questions concerning laterality, containment, and lobar or segmental location were answered correctly, whereas the three incorrect responses involved detailed relationships with surrounding airway or vascular structures. Participants also rated all three aspects highly, with spatial interpretation scoring 5.0. In Study 2, all participants successfully completed all three multi-step interaction scenarios. All three subject ratings exceeded 4.8. 

\subsection{Qualitative Feedback}
Overall, all six aspects received high ratings. Ease of learning and intuitiveness of natural-language interaction received the highest ratings, both at 5.0. The lowest rating was for intention to use MedVA for similar medical visualization tasks, although it remained high at 4.67. 

Open-ended responses clarified the aspects of MedVA that participants found useful. Novice participants particularly valued the ability to obtain ROI-focused visualizations through natural-language requests without requiring extensive experience in volume visualization or manual ROI segmentation preparation. They also appreciated that the requested ROIs could be complemented with relevant surrounding anatomical structures. The medical researchers similarly valued the automation of operations that would otherwise require programming or repeated manual adjustment of rendering parameters, and highlighted the benefits of visualizing pathological ROIs together with their relevant anatomical context. In particular, one researcher noted that visualizing the relationship between a lung tumor and surrounding pulmonary vessels could support surgical planning. Such visualization could clarify which vessels are adjacent to or closely associated with the tumor and help clinicians focus on them during intervention.

Response latency was the most consistent remaining interaction concern. Several participants reported that long response times interrupted the interaction flow and that faster responses would improve the experience. However, tolerance for this latency varied across participants; one novice participant noted that the delay was understandable given the use of an MLLM-based agentic system. 

\section{Discussions}
The results highlight the three complementary strengths of MedVA for medical volume visualization: (i) the neuro-symbolic intent formulation agent transforms potentially underspecified requests into clinically grounded explicit visualization specifications through symbolic reasoning (see \cref{tab:tab2}); (ii) the multi-model ROI identification agent identifies the semantically specified ROIs directly in the original volume through orchestration of pretrained segmentation models (see \cref{fig:fig3}); and (iii) the objective-driven visualization optimization agent translates the identified ROIs and their priorities into ROI-guided visualizations through an explicit volume-based visibility objective (see \cref{fig:fig4} and \cref{tab:tab4}). Together, these agents establish an end-to-end pathway from natural-language requests to volume visualizations that integrate established clinical knowledge with the underlying volumetric data (see \cref{fig:fig5,fig:fig6,fig:fig7}). The user study further supported the usability and practical value of this end-to-end workflow (Sections 7.3-7.4).  

Clinically meaningful medical volume visualization should convey not only relevant ROIs but also their relationships and relative visualization priorities. In an agentic setting, formulating such visualization goals from potentially underspecified requests requires both semantic flexibility and reliable grounding in established clinical knowledge. Prior systems largely rely on the broad medical knowledge encoded in MLLMs and their general reasoning capabilities to interpret user intent and formulate the corresponding visualization goals. Although this provides flexible semantic reasoning, these systems lack explicit mechanisms for grounding the intent formulation process in established clinical knowledge. This limitation was reflected in our results (see \cref{tab:tab2}), where the MLLM-only baseline left broad clinical concepts unresolved or omitted relevant context ROIs. Our neuro-symbolic intent formulation agent instead leverages symbolic reasoning over explicit clinical knowledge to validate and refine MLLM-derived visualization goals. The resulting visualization specification retains the flexibility of natural-language interaction while providing an explicit and verifiable reference for downstream ROI identification and visualization optimization. The practical value of this formulation process was supported by high intent alignment ratings in Study 1 (\cref{tab:tab5}) and novice feedback highlighting the benefits of complementing requested ROIs with relevant anatomical context. 

For agentic medical volume visualization systems, semantically specified ROIs need to be associated with the original medical volume. This association should also be available on demand to accommodate varying combinations of anatomical and pathological ROIs. Prior agentic systems are limited in both aspects because MLLMs often infer ROIs indirectly from multi-view renderings. Such indirect identification can be unreliable, particularly when multiple ROIs overlap in the rendered views or exhibit similar visual characteristics. Reliance on precomputed renderings limits on-demand identification because newly requested ROIs may require additional preprocessing. Our multi-model ROI identification agent addresses these limitations by invoking large-scale pretrained medical segmentation models according to the explicit visualization specification. This allows the semantically specified ROIs to be identified directly in the original medical volume at request time. The practical value of this capability was supported by the high ROI inclusion ratings in Study 1 and positive novice feedback on ROI-guided visualizations without manual segmentation preparation.

Two large-scale pretrained segmentation models achieved strong overall segmentation performance, with varying accuracy across ROIs (see \cref{tab:tab3}). Our multi-model ROI identification agent therefore evaluates complementary candidate segmentations from both models using the deterministic selection and MLLM-based validation modules and combines the appropriate model-specific outputs. This orchestration broadened ROI coverage while providing higher-quality segmentations for individual ROIs (see \cref{fig:fig3}).

At the visualization optimization stage, the identified ROIs and visualization priorities need to be translated into rendering parameters that emphasize the primary ROIs while preserving relevant context ROIs. Prior agentic systems are limited in this aspect because they largely rely on rendering-based VPA loops, in which MLLMs implicitly evaluate intermediate rendering outputs and adjust rendering parameters. Such implicit inference can be unreliable when optimization must account for complex visibility and occlusion relationships among multiple ROIs, as commonly encountered in medical volume visualization. Our objective-driven visualization optimization agent instead computes ROI visibility explicitly in the original volume using the identified ROIs and visualization specification. This volume-based visibility objective produced ROI-guided visualizations that reflected the intended visual emphasis (\cref{fig:fig4}). The practical effectiveness of this capability was supported by the high spatial interpretation rating in Study 1 (see \cref{tab:tab5}). Medical researcher feedback highlighted the benefits of automating repeated rendering parameter adjustments. The computational results showed that the explicit volume-based visibility objective substantially improved efficiency by reducing the number of iterations (see \cref{tab:tab4}). 

Effective end-to-end agentic systems require a coherent pathway from user requests to visualizations. Prior agentic systems remain limited in this aspect because they rely heavily on inference-driven decision making across stages, and uncertainty at one stage may propagate to subsequent stages. In contrast, MedVA establishes an explicit pathway that carries the user intent through visualization specification, volume-based ROI identification, and objective-driven visualization optimization. This explicit pathway allows MedVA to adapt ROI selection and visual emphasis to user intents with different levels of specificity (see \cref{fig:fig5}). The results under inter-patient and inter-dataset variations also showed that MedVA consistently completed the entire workflow and produced comparable ROI-guided visualizations (see \cref{fig:fig8}). The practical effectiveness of this integrated workflow was supported by the successful completion of all Study 2 scenarios and the high ratings for request fulfillment and visualization preservation (\cref{tab:tab5}). 

Agentic systems for medical volume visualization require diverse forms of interaction and different rendering modes. The multi-step interaction results showed that MedVA supported progressive ROI addition, TF manipulation, rendering mode changes, and visualization querying while preserving the previous visualization context (see \cref{fig:fig6,fig:fig7}). The user experience ratings indicated high usability of these capabilities, with particularly high ratings for ease of learning and the intuitiveness of natural-language interaction (see \cref{tab:tab5}). Participants also reported a high intention to use MedVA for similar medical volume visualization tasks. Although the end-to-end runtime remained relatively long (see \cref{fig:fig9}), the overall user ratings suggest that this latency did not preclude practical use in the evaluated setting.

\section{Conclusion}
We introduced MedVA, an end-to-end neuro-symbolic agentic system for medical volume visualization. MedVA complements MLLM-based reasoning with explicit clinical knowledge and connects natural-language requests to clinically grounded volume visualizations through three complementary agents for intent formulation, ROI identification, and visualization optimization. The neuro-symbolic intent formulation agent converts potentially underspecified requests into explicit visualization specifications, while the multi-model ROI identification agent directly identifies the specified targets in the original volume through pretrained segmentation models. The objective-driven visualization optimization agent then uses explicit ROI visibility to optimize visual emphasis and occlusion according to the visualization goal.

MedVA also has several limitations and opportunities for future work. First, MedVA depends on the coverage of available clinical knowledge and segmentation models. Although these resources adequately covered the evaluated cases, broader settings may include anatomical or pathological concepts, clinical relationships, or ROIs beyond this coverage. Future work will incorporate additional clinical knowledge sources and anatomy- or pathology-specific segmentation models. Second, although MedVA supported all evaluated workflows, wider use will require greater end-to-end efficiency. We will explore faster inference, reuse of target and scene information, and greater workflow parallelization. Finally, although our evaluation covered multiple datasets, scenarios, and users, broader studies are needed to establish generalizability across more diverse cases, clinicians, and medical researchers.

\bibliographystyle{abbrv-doi-hyperref}

\bibliography{template}

@Article{Zhang2011VolumeVisualizationTechnical,
author  = {Zhang, Qi and Eagleson, Roy and Peters, Terry M.},
journal = {J Digit Imaging},
title   = {Volume visualization: A technical overview with a focus on medical applications},
year    = {2011},
month   = aug,
number  = {4},
pages   = {640--664},
volume  = {24},
longdoi = {10.1007/s10278-010-9321-6},
}

@Article{Zhou2022ReviewThreeDimensional,
author    = {Zhou, Liang and Fan, Mengjie and Hansen, Charles and Johnson, Chris R. and Weiskopf, Daniel},
journal   = {Health Data Sci},
title     = {A review of three-dimensional medical image visualization},
year      = {2022},
month     = apr,
articleno = {9840519},
volume    = {2022},
numpages  = {19},
longdoi   = {10.34133/2022/9840519},
}

@Article{Svakhine2009IllustrationInspiredDepth,
author  = {Svakhine, Nikolai A. and Ebert, David S. and Andrews, William M.},
journal = {IEEE Trans Vis Comput Graph},
title   = {Illustration-inspired depth enhanced volumetric medical visualization},
year    = {2009},
month   = jan,
number  = {1},
pages   = {77--86},
volume  = {15},
longdoi = {10.1109/TVCG.2008.56},
}

@InProceedings{Viola2004ImportanceDrivenVolume,
author    = {Viola, Ivan and Kanitsar, Armin and Gr{\"o}ller, Meister Eduard},
booktitle = {Proc.\ Visualization},
title     = {Importance-driven volume rendering},
year      = {2004},
month     = oct,
pages     = {139--145},
series    = {VIS},
longdoi   = {10.1109/VISUAL.2004.48},
}

@Article{Li2025AttentionDrivenVisual,
author  = {Li, Mingjian and Jung, Younhyun and Song, Shaoli and Kim, Jinman},
journal = {Visual Comput},
title   = {Attention-driven visual emphasis for medical volumetric image visualization},
year    = {2025},
number  = {5},
pages   = {3205--3219},
volume  = {41},
longdoi = {10.1007/s00371-024-03596-9},
}

@Article{Cai2013AutomaticTransferFunction,
author  = {Cai, Lile and Tay, Wei-Liang and Nguyen, Binh P. and Chui, Chee-Kong and Ong, Sim-Heng},
journal = {Comput Med Imaging Graph},
title   = {Automatic transfer function design for medical visualization using visibility distributions and projective color mapping},
year    = {2013},
month   = oct,
number  = {7--8},
pages   = {450--458},
volume  = {37},
longdoi = {10.1016/j.compmedimag.2013.08.008},
}

@Article{RezkSalama2006HighLevelUser,
author  = {Rezk-Salama, Christof and Keller, Maik and Kohlmann, Peter},
journal = {IEEE Trans Vis Comput Graph},
title   = {High-level user interfaces for transfer function design with semantics},
year    = {2006},
month   = sep,
number  = {5},
pages   = {1021--1028},
volume  = {12},
longdoi = {10.1109/TVCG.2006.148},
}

@Article{Liu2024AVATowardsAutonomous,
author  = {Liu, Shusen and Miao, Hanqi and Li, Zhanping and Olson, Matthew and Pascucci, Valerio and Bremer, Peer-Timo},
journal = {Comput Graph Forum},
title   = {{AVA}: Towards autonomous visualization agents through visual perception-driven decision-making},
year    = {2024},
month   = jun,
number  = {3},
articleno = {e15093},
volume  = {43},
longdoi = {10.1111/cgf.15093},
}

@Article{Ai2026NLI4VolVisNaturalLanguage,
author  = {Ai, Kuangshi and Tang, Kaiyuan and Wang, Chaoli},
journal = {IEEE Trans Vis Comput Graph},
title   = {{NLI4VolVis}: Natural language interaction for volume visualization via {LLM} multi-agents and editable {3D} Gaussian splatting},
year    = {2026},
month   = jan,
number  = {1},
pages   = {46--56},
volume  = {32},
longdoi = {10.1109/TVCG.2025.3633888},
}

@InProceedings{Mallick2024ChatVisAutomatingScientific,
author    = {Mallick, Tanwi and Yildiz, Orcun and Lenz, David and Peterka, Tom},
booktitle = {Proc.\ ACM/IEEE SC Workshops},
title     = {{ChatVis}: Automating scientific visualization with a large language model},
year      = {2024},
pages     = {49--55},
longdoi   = {10.1109/SCW63240.2024.00014},
}

@Misc{Wang2025IntuiTFMLLMGuided,
author       = {Wang, Yiyao and Pan, Bo and Wang, Ke and Liu, Han and Mao, Jinyuan and Liu, Yuxin and Zhu, Minfeng and Huang, Xiuqi and Chen, Weifeng and Zhang, Bo and Chen, Wei},
title        = {{IntuiTF}: {MLLM}-guided transfer function optimization for direct volume rendering},
year         = {2025},
howpublished = {arXiv preprint arXiv:2506.18407},
longdoi      = {10.48550/arXiv.2506.18407},
}

@Article{Biswas2026VizGenieToward,
author  = {Biswas, Ayan and Turton, Terece L. and Ranasinghe, Nishath Rajiv and Jones, Shawn and Love, Bradley and Jones, William and Hagberg, Aric and Shen, Han-Wei and DeBardeleben, Nathan and Lawrence, Earl},
journal = {IEEE Trans Vis Comput Graph},
title   = {{VizGenie}: Toward self-refining, domain-aware workflows for next-generation scientific visualization},
year    = {2026},
month   = jan,
number  = {1},
pages   = {1021--1031},
volume  = {32},
longdoi = {10.1109/TVCG.2025.3634655},
}

@Misc{Sun2026SASAVSelfDirected,
author       = {Sun, Jianxin and Lenz, David and Peterka, Tom and Yu, Hongfeng},
title        = {{SASAV}: Self-directed agent for scientific analysis and visualization},
year         = {2026},
howpublished = {arXiv preprint arXiv:2604.03406},
longdoi      = {10.48550/arXiv.2604.03406},
}

@Misc{Ai2026HiLSVADesignEvaluation,
author       = {Ai, Kuangshi and Do, Patrick Phuoc and Wang, Chaoli},
title        = {{HiLSVA}: Design and evaluation of a human-in-the-loop agentic system for scientific visualization},
year         = {2026},
howpublished = {arXiv preprint arXiv:2606.26614},
longdoi      = {10.48550/arXiv.2606.26614},
}

@Article{Hitzler2022NeuroSymbolicApproaches,
author  = {Hitzler, Pascal and Eberhart, Aaron and Ebrahimi, Monireh and Sarker, Md Kamruzzaman and Zhou, Lu},
journal = {Natl Sci Rev},
title   = {Neuro-symbolic approaches in artificial intelligence},
year    = {2022},
month   = jun,
number  = {6},
articleno = {nwac035},
volume  = {9},
longdoi = {10.1093/nsr/nwac035},
}

@Article{Wang2025TowardsDataKnowledgeDriven,
author  = {Wang, Wenguan and Yang, Yi and Wu, Fei},
journal = {IEEE Trans Pattern Anal Mach Intell},
title   = {Towards data- and knowledge-driven {AI}: A survey on neuro-symbolic computing},
year    = {2025},
month   = feb,
number  = {2},
pages   = {878--899},
volume  = {47},
longdoi = {10.1109/TPAMI.2024.3483273},
}

@Article{Hakim2026NeuroSymbolicAgentic,
author  = {Hakim, Safayat Bin and Adil, Muhammad and Velasquez, Alvaro and Song, Houbing Herbert},
journal = {Comput Sci Rev},
title   = {Neuro-symbolic agentic {AI}: Architectures, integration patterns, applications, open challenges and future research directions},
year    = {2026},
month   = may,
articleno = {100902},
volume  = {60},
longdoi = {10.1016/j.cosrev.2026.100902},
}

@InProceedings{Jia2025medIKALIntegratingKnowledge,
author    = {Jia, Mingyi and Duan, Junwen and Song, Yan and Wang, Jianxin},
booktitle = {Proc.\ COLING},
title     = {med{IKAL}: Integrating knowledge graphs as assistants of {LLM}s for enhanced clinical diagnosis on {EMR}s},
year      = {2025},
month     = jan,
pages     = {9278--9298},
address   = {Abu Dhabi, UAE},
publisher = {Association for Computational Linguistics},
}

@Article{Prenosil2025NeuroSymbolicAIAuditable,
author  = {Prenosil, George A. and Weitzel, Thilo K. and Bello, Sandra C. and Mingels, Clemens and Manzini, Giulia and Meier, Lorenz P. and Shi, Kuang-Yu and Rominger, Axel and Afshar-Oromieh, Ali},
journal = {Commun Med},
title   = {Neuro-symbolic {AI} for auditable cognitive information extraction from medical reports},
year    = {2025},
month   = nov,
articleno = {491},
volume  = {5},
longdoi = {10.1038/s43856-025-01194-x},
}

@Article{Wasserthal2023TotalSegmentatorRobust,
author    = {Wasserthal, Jakob and Breit, Hanns-Christian and Meyer, Manfred T. and Pradella, Maurice and Hinck, Daniel and Sauter, Alexander W. and Heye, Tobias and Boll, Daniel T. and Cyriac, Joshy and Yang, Shan and Bach, Michael and Segeroth, Martin},
journal   = {Radiol Artif Intell},
title     = {{TotalSegmentator}: Robust segmentation of 104 anatomic structures in {CT} images},
year      = {2023},
month     = sep,
number    = {5},
articleno = {e230024},
volume    = {5},
longdoi   = {10.1148/ryai.230024},
}

@InProceedings{He2025VISTA3DUnifiedSegmentation,
author    = {He, Yufan and Guo, Pengfei and Tang, Yucheng and Myronenko, Andriy and Nath, Vishwesh and Xu, Ziyue and Yang, Dong and Zhao, Can and Simon, Benjamin and Belue, Mason and Harmon, Stephanie and Turkbey, Baris and Xu, Daguang and Li, Wenqi},
booktitle = {Proc.\ CVPR},
title     = {{VISTA3D}: A unified segmentation foundation model for {3D} medical imaging},
year      = {2025},
month     = jun,
pages     = {20863--20873},
}

@InProceedings{Rokuss2026VoxTellFreeText,
author    = {Rokuss, Maximilian and Langenberg, Moritz and Kirchhoff, Yannick and Isensee, Fabian and Hamm, Benjamin and Ulrich, Constantin and Regnery, Sebastian and Bauer, Lukas and Katsigiannopulos, Efthimios and Norajitra, Tobias and Maier-Hein, Klaus},
booktitle = {Proc.\ CVPR},
title     = {{VoxTell}: Free-text promptable universal {3D} medical image segmentation},
year      = {2026},
month     = jun,
pages     = {37538--37557},
}

@InProceedings{Schubert1993SpatialKnowledgeRepresentation,
author    = {Schubert, Rainer and H{\"o}hne, Karl Heinz and Pommert, Andreas and Riemer, Martin and Schiemann, Thomas and Tiede, Ulf},
booktitle = {Proc.\ IPMI},
title     = {Spatial knowledge representation for visualization of human anatomy and function},
year      = {1993},
month     = jun,
pages     = {168--181},
volume    = {687},
series    = {Lect Notes Comput Sci},
publisher = {Springer},
longdoi   = {10.1007/BFb0013787},
}

@InProceedings{Pommert1994SymbolicModelingHuman,
author    = {Pommert, Andreas and Schubert, Rainer and Riemer, Martin and Schiemann, Thomas and Tiede, Ulf and H{\"o}hne, Karl Heinz},
booktitle = {Proc.\ Visualization in Biomedical Computing},
title     = {Symbolic modeling of human anatomy for visualization and simulation},
year      = {1994},
month     = sep,
pages     = {412--423},
volume    = {2359},
series    = {Proc SPIE},
publisher = {SPIE},
longdoi   = {10.1117/12.185202},
}

@InProceedings{Wong1999SemiAutomaticSceneGeneration,
author    = {Wong, B. A. and Rosse, Cornelius and Brinkley, James F.},
booktitle = {Proc.\ AMIA Symp},
title     = {Semi-automatic scene generation using the {Digital Anatomist Foundational Model}},
year      = {1999},
pages     = {637--641},
}

@Article{Ezquerra1999InteractiveKnowledgeGuided,
author  = {Ezquerra, Norberto F. and de Braal, Levien and Garcia, Ernest V. and Cooke, C. David and Krawczynska, Elizabeth},
journal = {Future Gener Comput Syst},
title   = {Interactive, knowledge-guided visualization of {3D} medical imagery},
year    = {1999},
month   = feb,
number  = {1},
pages   = {59--73},
volume  = {15},
longdoi = {10.1016/S0167-739X(98)00055-7},
}

@Article{Correa2011VisibilityHistograms,
author  = {Correa, Carlos D. and Ma, Kwan-Liu},
journal = {IEEE Trans Vis Comput Graph},
title   = {Visibility histograms and visibility-driven transfer functions},
year    = {2011},
month   = feb,
number  = {2},
pages   = {192--204},
volume  = {17},
longdoi = {10.1109/TVCG.2010.35},
}

@Article{Bruckner2006IllustrativeContextPreserving,
author  = {Bruckner, Stefan and Grimm, S{\"o}ren and Kanitsar, Armin and Gr{\"o}ller, Meister Eduard},
journal = {IEEE Trans Vis Comput Graph},
title   = {Illustrative context-preserving exploration of volume data},
year    = {2006},
month   = nov,
number  = {6},
pages   = {1559--1569},
volume  = {12},
longdoi = {10.1109/TVCG.2006.96},
}

@InProceedings{Kambhampati2024LLMsCantPlan,
author    = {Kambhampati, Subbarao and Valmeekam, Karthik and Guan, Lin and Verma, Mudit and Stechly, Kaya and Bhambri, Siddhant and Saldyt, Lucas and Murthy, Anil},
booktitle = {Proc.\ ICML},
title     = {Position: {LLM}s can't plan, but can help planning in {LLM}-modulo frameworks},
year      = {2024},
pages     = {22895--22907},
volume    = {235},
series    = {Proc Mach Learn Res},
}

@InProceedings{Pan2023LogicLMEmpowering,
author    = {Pan, Liangming and Albalak, Alon and Wang, Xinyi and Wang, William Yang},
booktitle = {Findings of EMNLP},
title     = {Logic-{LM}: Empowering large language models with symbolic solvers for faithful logical reasoning},
year      = {2023},
month     = dec,
pages     = {3806--3824},
address   = {Singapore},
publisher = {Association for Computational Linguistics},
longdoi   = {10.18653/v1/2023.findings-emnlp.248},
}

@InProceedings{Sun2024ThinkOnGraphDeep,
author    = {Sun, Jiashuo and Xu, Chengjin and Tang, Lumingyuan and Wang, Saizhuo and Lin, Chen and Gong, Yeyun and Ni, Lionel and Shum, Heung-Yeung and Guo, Jian},
booktitle = {Proc.\ ICLR},
title     = {Think-on-Graph: Deep and responsible reasoning of large language model on knowledge graph},
year      = {2024},
}

@InProceedings{Stutz2025DIVENeuroSymbolic,
author    = {St{\"u}tz, Jan-David and Gegziabher, Selamawit and Blaga, Victor and Karras, Oliver and Oelen, Allard and Auer, S{\"o}ren},
booktitle = {Proc.\ WISE},
title     = {{DIVE}: A neuro-symbolic and user-centered approach for data insight visualization},
year      = {2025},
pages     = {433--443},
publisher = {Springer},
longdoi   = {10.1007/978-981-95-7251-9_29},
}

@InProceedings{Codiglione2025NL2SPARQLOntologyBased,
author    = {Codiglione, Matteo and Remondino, Fabio},
booktitle = {Proc.\ Digital Heritage},
title     = {{NL-2-SPARQL}: Ontology-based natural language querying over {3D} point cloud knowledge graphs},
year      = {2025},
publisher = {The Eurographics Association},
longdoi   = {10.2312/dh.20253280},
}

@Article{Rosse2003ReferenceOntologyBiomedical,
author  = {Rosse, Cornelius and Mejino, Jos{\'e} L. V.},
journal = {J Biomed Inform},
title   = {A reference ontology for biomedical informatics: The {Foundational Model of Anatomy}},
year    = {2003},
month   = dec,
number  = {6},
pages   = {478--500},
volume  = {36},
longdoi = {10.1016/j.jbi.2003.11.007},
}

@Article{Sioutos2007NCIThesaurusSemantic,
author  = {Sioutos, Nicholas and de Coronado, Sherri and Haber, Margaret W. and Hartel, Frank W. and Shaiu, Wen-Ling and Wright, Lawrence W.},
journal = {J Biomed Inform},
title   = {{NCI Thesaurus}: A semantic model integrating cancer-related clinical and molecular information},
year    = {2007},
month   = feb,
number  = {1},
pages   = {30--43},
volume  = {40},
longdoi = {10.1016/j.jbi.2006.02.013},
}

@Article{Zhuang2019EvaluationAlgorithmsMultiModality,
author    = {Zhuang, Xiahai and Li, Lei and Payer, Christian and {\v{S}}tern, Darko and Urschler, Martin and Heinrich, Mattias P. and Oster, Julien and Wang, Chunliang and Smedby, {\"O}rjan and Bian, Cheng and Yang, Xin and Heng, Pheng-Ann and Mortazi, Aliasghar and Bagci, Ulas and Yang, Guanyu and Sun, Chenchen and Galisot, Gaetan and Ramel, Jean-Yves and Brouard, Thierry and Tong, Qianqian and Si, Weixin and Liao, Xiangyun and Zeng, Guodong and Shi, Zenglin and Zheng, Guoyan and Wang, Chengjia and MacGillivray, Tom and Newby, David and Rhode, Kawal and Ourselin, Sebastien and Mohiaddin, Raad and Keegan, Jennifer and Firmin, David and Yang, Guang},
journal   = {Med Image Anal},
title     = {Evaluation of algorithms for multi-modality whole heart segmentation: An open-access grand challenge},
year      = {2019},
month     = dec,
articleno = {101537},
volume    = {58},
longdoi   = {10.1016/j.media.2019.101537},
}

@Article{Sang2026BenchmarkSegmentationPelvic,
author  = {Sang, Yudi and Liu, Yanzhen and Yibulayimu, Sutuke and Wang, Yunning and Killeen, Benjamin D. and Liu, Mingxu and Ku, Ping-Cheng and Johannsen, Ole and Gotkowski, Karol and Zenk, Maximilian and Maier-Hein, Klaus and Isensee, Fabian and Yue, Peiyan and Wang, Yi and Yu, Haidong and Pan, Zhaohong and He, Yutong and Liang, Xiaokun and Liu, Daiqi and Fan, Fuxin and Jurgas, Artur and Skalski, Andrzej and Ma, Yuxi and Yang, Jing and P{\l}otka, Szymon and Litka, Rafa{\l} and Zhu, Gang and Song, Yingchun and Unberath, Mathias and Armand, Mehran and Ruan, Dan and Zhou, S. Kevin and Cao, Qiyong and Zhao, Chunpeng and Wu, Xinbao and Wang, Yu},
journal = {IEEE Trans Med Imaging},
title   = {Benchmark of segmentation techniques for pelvic fracture in {CT} and {X}-ray: Summary of the {PENGWIN} 2024 challenge},
year    = {2026},
number  = {5},
pages   = {2212--2228},
volume  = {45},
}

@Article{Ramakrishnan2024LargeOpenAccess,
author  = {Ramakrishnan, Divya and Jekel, Lauren and Chadha, Shagun and Janas, Anna and Moy, Hannah and Maleki, Neda and Sala, Maria and Kaur, Manpreet and Petersen, Gregory C. and Merkaj, Sara and von Reppert, Michael and Baid, Ujjwal and Bakas, Spyridon and Kirsch, Claudia and Davis, Matthew and Bousabarah, Khaled and Holler, Wolfgang and Lin, Ming and Westerhoff, Michael and Aneja, Sanjay and Memon, Faisal and Aboian, Mariam S.},
journal = {Sci Data},
title   = {A large open access dataset of brain metastasis {3D} segmentations on {MRI} with clinical and imaging information},
year    = {2024},
volume  = {11},
longdoi = {10.1038/s41597-024-03021-9},
}

@Article{Reddy2026UniversityTexasSouthwestern,
author  = {Reddy, Divya D. and Saadat, Niloufar and Holcomb, James M. and Wagner, Benjamin C. and Truong, Nghi C. and Bowerman, Jason and Hatanpaa, Kimmo J. and Patel, Toral R. and Pinho, Marco C. and Yu, Fang and Zhang, Kuan and Lodhi, Sadeem and Madhuranthakam, Ananth J. and Yogananda, Chandan Ganesh Bangalore and Maldjian, Joseph A.},
journal = {Sci Data},
title   = {The University of Texas Southwestern glioma dataset -- {MRI}, molecular markers and segmentations},
year    = {2026},
month   = apr,
articleno = {934},
volume  = {13},
longdoi = {10.1038/s41597-026-07274-4},
}

@Misc{Baid2021RSNAASNRMICCAIBraTS,
author       = {Baid, Ujjwal and Ghodasara, Satyam and Mohan, Suyash and Bilello, Michel and Calabrese, Evan and Colak, Errol and Farahani, Keyvan and Kalpathy-Cramer, Jayashree and Kitamura, Felipe C. and Pati, Sarthak and Prevedello, Luciano M. and Rudie, Jeffrey D. and Sako, Chiharu and Shinohara, Russell T. and Bergquist, Timothy and Chai, Rong and Eddy, Jeffrey and Elliott, James and Reade, Walter and Schaffter, Thomas and Yu, Thomas and Zheng, Jun and Moawad, Ahmed W. and Coelho, La{\'e}rcio O. and McDonnell, Orla and Others},
title        = {The {RSNA-ASNR-MICCAI} {BraTS} 2021 benchmark on brain tumor segmentation and radiogenomic classification},
year         = {2021},
howpublished = {arXiv preprint arXiv:2107.02314},
}

@Article{Ferrara2026SharingWholeTotalBody,
author  = {Ferrara, D. and Pires, M. and Gutschmayer, S. and Yu, J. and Abdelhafez, Y. G. and Abenavoli, E. and Badawi, R. D. and Chaudhari, A. J. and Chen, M. S. and Cherry, S. R. and Frille, A. and Geist, B. K. and Gr{\"u}nert, S. and Hacker, M. and Hesse, S. and Kerkhoff, T. and Linder, P. and Pappisch, J. and Pusitz, S. and Raslan, O. A. and Rausch, I. and Raychaudhuri, S. P. and Sabri, O. and Schmidt, F. and Sciagr{\`a}, R. and Spencer, B. and Wang, G. and Wirtz, H. and Beyer, T. and Sundar, L. K. S.},
title   = {Sharing a whole-/total-body [{18F}]{FDG}-{PET}/{CT} dataset with {CT}-derived segmentations: An {ENHANCE.PET} initiative},
year    = {2026},
}

@Misc{IRCAD3DIRCADb01,
author       = {{IRCAD}},
title        = {{3D-IRCADb-01}: {3D} image reconstruction for comparison of algorithm database},
howpublished = {IRCAD Research Institute},
}

\end{document}